\documentclass[a4paper,12pt]{article}
\pdfoutput=1
\usepackage[T1]{fontenc}
\usepackage[utf8]{inputenc}
\usepackage{lmodern}
\usepackage[english]{babel}
\usepackage{xspace}

\usepackage{geometry} 
\usepackage{amsmath,amssymb,amsfonts,amsthm}

\usepackage{siunitx}
\DeclareSIUnit\particle{particle} 
\DeclareSIUnit\weight{wt \%}
\DeclareSIUnit\stiffness{\pico\newton/\micro\meter}
\DeclareSIUnit\kbT{k_{\text{B}} T}

\usepackage{graphicx}
\usepackage{caption}
\usepackage{subcaption} 

\usepackage{booktabs}
\usepackage{array}
\usepackage{paralist} 

\usepackage{textcomp} 
\usepackage{xcolor} 
\usepackage{cite} 

\usepackage{hyperref}
\usepackage{bookmark}

\usepackage[nottoc]{tocbibind} 
\usepackage{tocloft}

\definecolor{linkcolor}{rgb}{0,0,0.6} 
\hypersetup{
	colorlinks=true, 
	pdfstartview=FitV, 
	urlcolor=linkcolor, 
	linkcolor= linkcolor, 
	citecolor=linkcolor 
}

\begin{document}
\title{Fluctuations in an aging system: absence of effective temperature in the sol-gel transition of a quenched gelatin sample}
\author{Antoine B\'{e}rut$^{1}$, Artyom Petrosyan$^{1}$, \\ Juan Ruben Gomez-Solano$^{2}$  and Sergio Ciliberto$^{1}$ \\
\\ \\
$^{1}$ Universit\'{e} de Lyon, \\ \'{E}cole Normale Sup\'{e}rieure de Lyon, \\
Laboratoire de Physique, C.N.R.S. UMR5672, \\ 46, All\'{e}e d'Italie, 69364 Lyon, France \\
$^{2}$ 2. Physikalisches Institut, Universit\"{a}t Stuttgart, \\ Pfaffenwaldring 57, 70569 Stuttgart, Germany}
\maketitle

\begin{abstract}
We study the fluctuations of a Brownian micro particle trapped with optical tweezers in a gelatin solution undergoing a fast local temperature quench below the sol-gel transition. Contrary to what was previously reported, we observe no anomalous fluctuations in the particle's position that could be interpreted in terms of an effective temperature. A careful analysis with ensemble averages shows only equilibrium-like properties for the fluctuations, even though the system is clearly aging. We also provide a detailed discussion on possible artifacts that could have been interpreted as an effective temperature, such as the presence of a drift or a mixing in time and ensemble averages in data analysis. These considerations are of general interest when dealing with non-ergodic or non-stationary systems.
\end{abstract}

\tableofcontents

\section{Introduction and Motivations}

\subsection{Gelatin and the sol-gel transition}

Gelatin is a thermoreversible gel~\cite{ThermoreversibleNetworks}. It is a heterogeneous mixture of water-soluble denatured collagen protein chains, extracted by boiling animal by-products (skin, tendons, ligaments, bones, etc.) in water. Collagen molecules are rods of \SI{300}{\nano\meter} length, made of three strands, with high average molecular weights. This triple-helix structure is stabilised by hydrogen bonds and has a diameter of $\sim \SI{1.4}{\nano\meter}$. The chemical treatment used to produce gelatin breaks crosslinks between strands, but can also hydrolyze strands into fragments. Thus a broad molecular weight distribution is obtained for gelatin~\cite{djabourov1988gelationI,djabourov1988gelationII}.

Above a temperature $T_{\text{melt}} \sim \SI{40}{\celsius}$, the gelatin chains are in coil conformation. The gelatin solution is in a viscous liquid phase, called ``sol'' phase. Below a temperature $T_{\text{gel}} \sim \SI{30}{\celsius}$ renaturation
of the native triple helix structure occurs, and chains form a percolating three-dimensional network of helical segments connected by single strand coils. The gelatin solution is in an arrested state with elastic behaviur, called ``gel'' phase. The coil-helix transition is completely reversible and the transition from one phase to the other is called the ``sol-gel'' transition~\cite{ThermoreversibleNetworks}.

Physical properties of the sol phase, and of the sol-gel transition are studied in~\cite{Sidharta1993,Sidharta1994} for different gelatin concentrations above \SI{4}{\weight}. In particular, it was seen that there are at least three successive steps in the transition: monomer to aggregate formation, random-coil-single-helix transition (disorder-order transition), and single-helix-triple-helix transition (order-order transition). It is then possible to identify different phase states in the sol domain: the sol state I where the chains have random coil conformations and the sol state II where single and triple helices begin to form (without reaching gelation).

The gel phase was also shown to share properties with glassy materials, which are out-of-equilibrium metastable systems. After a quench at $T < T_{\text{gel}}$ the system is frustrated by topological constraints because each gelatin chain is involved in at least two helices, and neighbouring helices are competing for the shared portions of non-helical chain. Therefore, the system displays physical ageing: its physical properties slowly evolve with time, through a process known as structural recovery. For example, the small-strain shear modulus of a \SI{5}{\weight} gelatin solution quenched at \SI{20}{\celsius} increases logarithmically as a function of the ageing time~\cite{Caroli2009}. And the elasticity of gelatin gels during slow cool and heat cycles exhibits memory and rejuvenation effects similar to the ones found in spin glasses~\cite{Normand2010}.

Although it is known that mechanical properties of gelatin gels are very sensitive to temperature variations, previous thermal history of the gel, and time, this system has some interesting experimental features:
\begin{itemize}
\item The fact that the transition is thermoreversible allows us to do melting/gelation cycles simply by controlling the temperature of the sample.
\item The ageing rate can in theory be controlled by changing the quench depth.
\item The length-scale of the collagen chains (\SI{300}{\nano\meter}) is big enough to be sensed by a micro-particle of \SI{2}{\micro\meter}.
\end{itemize}
This particular sol-gel transition was chosen for previous works done in our group about fluctuations of Brownian particles in quenched gelatin samples~\cite{TheseRuben,RubenPRL2011,RubenEPL2012}.

A summary of the previous works results and our motivations are presented in the next section.

\subsection{Previous work: anomalous variance, heat flux and Fluctuation Dissipation Theorem violation in an ageing bath}

Previous works~\cite{TheseRuben,RubenPRL2011,RubenEPL2012} showed that a particle trapped with optical tweezers in a liquid droplet of gelatin solution, quenched at a temperature below $T_{\text{gel}}$ exhibits anomalously high position fluctuations right after the quench. These anomalous fluctuations could be interpreted as an effective temperature, and were consistent with a violation of the Fluctuation Dissipation Theorem (FDT) and with an exchange Fluctuation Theorem (xFT) for the heat exchanged between two heat baths at different temperatures~\cite{Jarzynski2004}.

\begin{figure}[ht!]
        \centering
        \begin{subfigure}[b]{0.5\textwidth}
                \includegraphics[width=\textwidth]{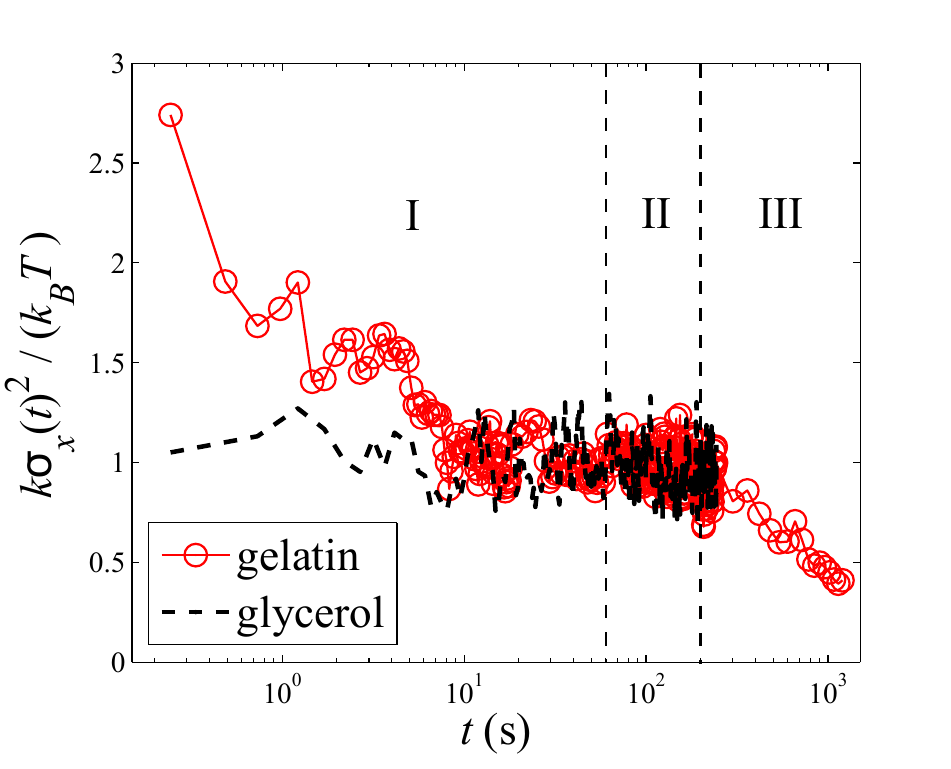}
                \caption{ }
                \label{gel:fig:Ruben_1}
        \end{subfigure}%
        \begin{subfigure}[b]{0.5\textwidth}
                \includegraphics[width=\textwidth]{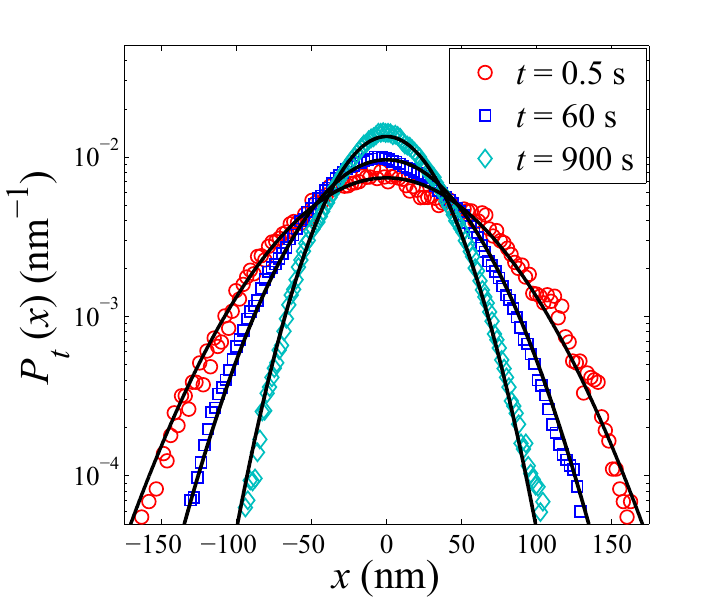}
                \caption{ }
                \label{gel:fig:Ruben_2}
        \end{subfigure}
        \caption{(a) Evolution of the normalised variance $k\sigma_{x}^2/k_{\text{B}}T$ of the position fluctuations of one particle trapped in gelatin solution (\SI{10}{\weight}) or glycreol, quenched at \SI{26}{\celsius}, for different times $t$ after the quench. (b) Evolution of the Probability Distribution Function of the position fluctuations of the particle trapped in gelatin solution.}\label{gel:fig:Ruben_1_et_2}
\end{figure}

\begin{figure}[ht!]
\begin{center}
\includegraphics[width=0.5\textwidth]{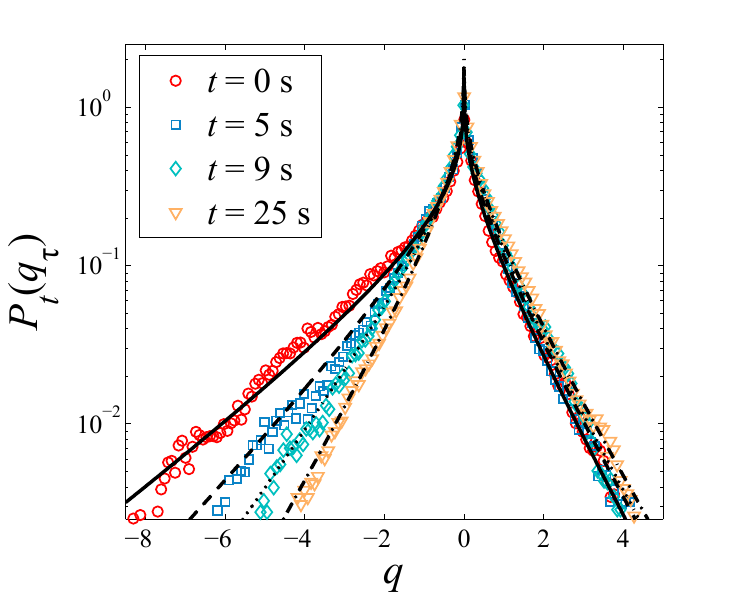}
\caption{Probability Density Function of the normalised heat $q$ exchanged during $\tau = \SI{30}{\second}$ computed at different times $t$ after the quench in gelatin.} \label{gel:fig:Ruben_3}
\end{center}
\end{figure}

\begin{figure}[ht!]
\begin{center}
\includegraphics[width=\textwidth]{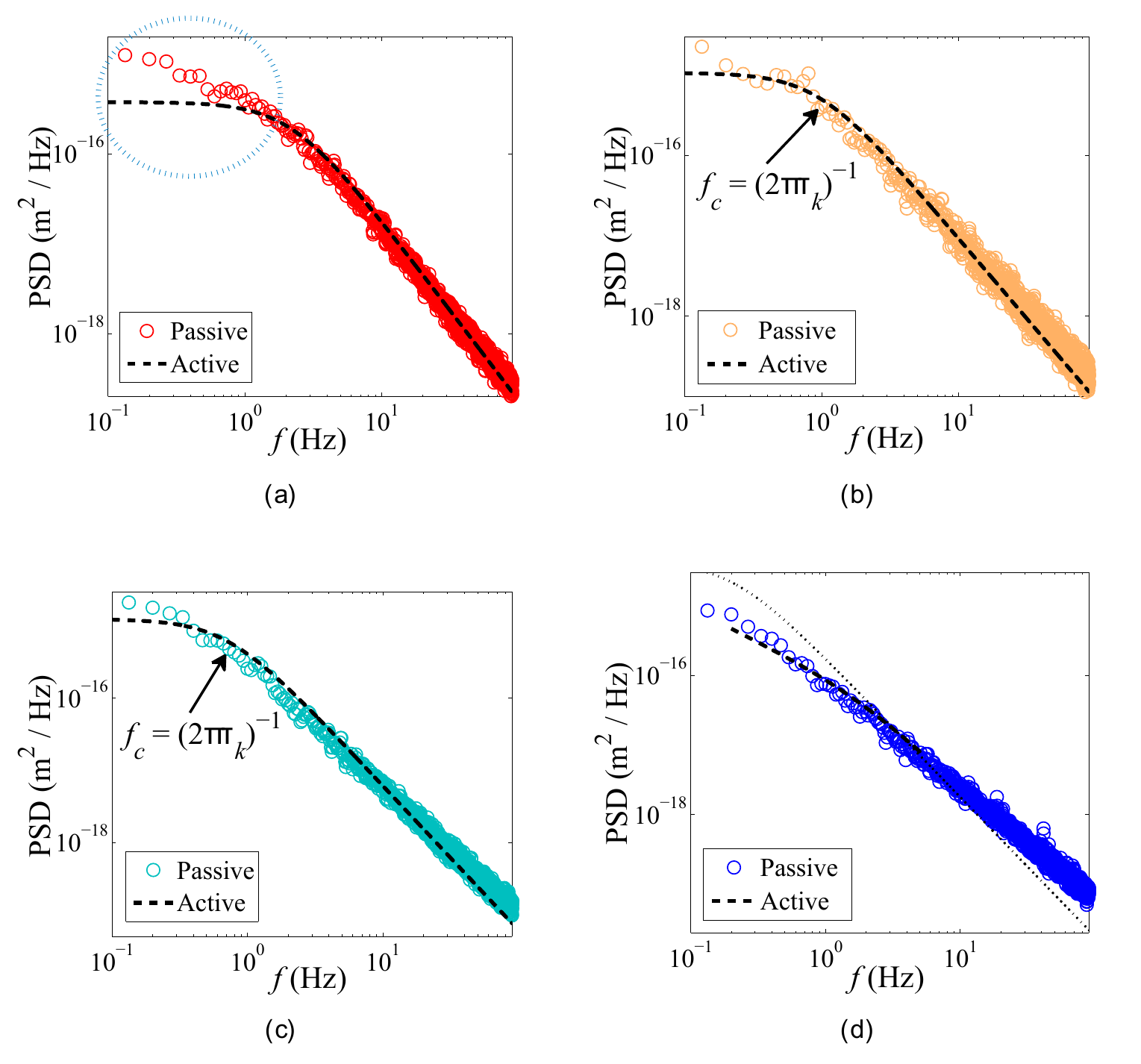}
\caption{Passive Power Spectral Densities of the position fluctuations (color points) and Fourier transform of the active response function (black dashed-lines) computed at different times after the quench. (a) For $ \SI{0}{\second} < t < \SI{15}{\second}$. (b) For $ \SI{30}{\second} < t < \SI{45}{\second}$. (c) For $ \SI{75}{\second} < t < \SI{90}{\second}$. (d) For $ \SI{1200}{\second} < t < \SI{1215}{\second}$. If the Fluctuations Dissipation Theorem is verified, the two quantities should be equal, which is not the case for low-frequency in (a).} \label{gel:fig:Ruben_4}
\end{center}
\end{figure}

We reproduce here some figures from~\cite{TheseRuben,RubenPRL2011,RubenEPL2012} and recall the associated key results:
\begin{itemize}
\item The variance of the position $\sigma_{x}^2 = \langle x^2 \rangle$ exhibits anomalously high value for short times ($\sim \SI{5}{\second}$) right after the quench. It then stabilises at the equipartition value $\sigma_{x \, \text{eq}}^2 = k_{\text{B}}T/k$ for $\sim \SI{200}{\second}$ (where $k_{\text{B}}$ is the Boltzmann constant and $k$ the stiffness of the trap). And it finally decreases logarithmically for long times after the quench. See figure \ref{gel:fig:Ruben_1}. The anomalously high variances can be interpreted in terms of effective temperatures: $\sigma_{x}^2 = k_{\text{B}}T_{\text{eff}}/k > \sigma_{x \, \text{eq}}^2$.
\item The Probability Distribution Functions of position fluctuations are Gaussian at any time after the quench, but their variances decreases with time (in agreement with the variances observed). See figure \ref{gel:fig:Ruben_2}.
\item The Probability Distribution Function of the heat exchanged between the particle and the bath during short times after the quench is asymmetrical. See figure \ref{gel:fig:Ruben_3}.
\item The asymmetry function of the heat $\rho (q) = \ln\left(P(q)/P(-q)\right)$ (where $P(q)$ is the probability of observing the value $q$ of the heat) satisfies an exchange Fluctuation Theorem: $\rho (q) = \Delta \beta q$, with its slope $\Delta \beta$ directly linked to the effective temperatures defined from the variances (see~\cite{RubenPRL2011}).
\item The Fluctuation Dissipation Theorem is violated only for short times after the quench, and this violation can be linked with the amount of heat exchanged between the particle and the bath during the same time. See figure \ref{gel:fig:Ruben_4}.
\end{itemize}

Unfortunately, none of those key results was found to be reproducible, and we believe that they were only due to an artifact in the data and/or in the analysis method. Therefore, we present in this article a detailed and careful analysis of trajectories of particles trapped in a droplet of gelatin solution quenched at a temperature below $T_{\text{gel}}$. We show that there is indeed no effective temperature for this system, which surprisingly exhibits equilibrium-like properties while aging.

\noindent The article is organized as follow: in the first section we describe our experimental set-up, the gelatin solution preparation, optical trapping system, and local heating for fast quenching method. In the second section, we discuss our experimental results: we show with some bulk measurements that the system is indeed aging after the quench, we then analyze the effect of a slow drift in trajectory and of a mixing between time and ensemble averages in data analysis to show that there is no anomalous variance of the particle's position. We end by discussing the consequences of the absence of effective temperature for the heat exchange, Fluctuation Theorem and Fluctuation Dissipation violation.

\newpage

\section{Experimental set-up}

\subsection{Gelatin sample preparation}
\label{gel:section:sample_preparation}

We use gelatin powder from porcine skin, produced by Sigma-Aldrich$^{\circledR}$: gel strength $\sim$~300~g Bloom, Type A, BioReagent, suitable for cell culture. This gelatin is derived from acid-cured tissue, whereas type B is derived from lime-cured tissue.


We work with gelatin at a weight concentration of \SI{5}{\weight}. The samples are prepared following a standard protocol~\cite{normand2000gelation}: the wanted amount of powder is dissolved in bidistilled water, which is then heated for $\sim \SI{30}{\minute}$ at $\sim \SI{60}{\celsius}$ while slowly stirred until the solution is transparent and homogeneous. While the solution is still liquid, $\sim \SI{2}{\milli\liter}$ are filtered using a Millex$^{\circledR}$ syringe driven filter unit with \SI{0.45}{\micro\meter} pore size mixed cellulose esters membrane. Then \SI{15}{\micro\liter} of an aqueous solution of silica beads (radius $R = \SI{1.00(5)}{\micro\meter}$) with concentration \SI{e7}{\particle\per\milli\liter} are added, and the solution is strongly agitated. The non-filtered and final solutions are let gel at room temperature and kept in the refrigerator for later use.

\begin{figure}[ht!]
\begin{center}
\includegraphics[width=12cm]{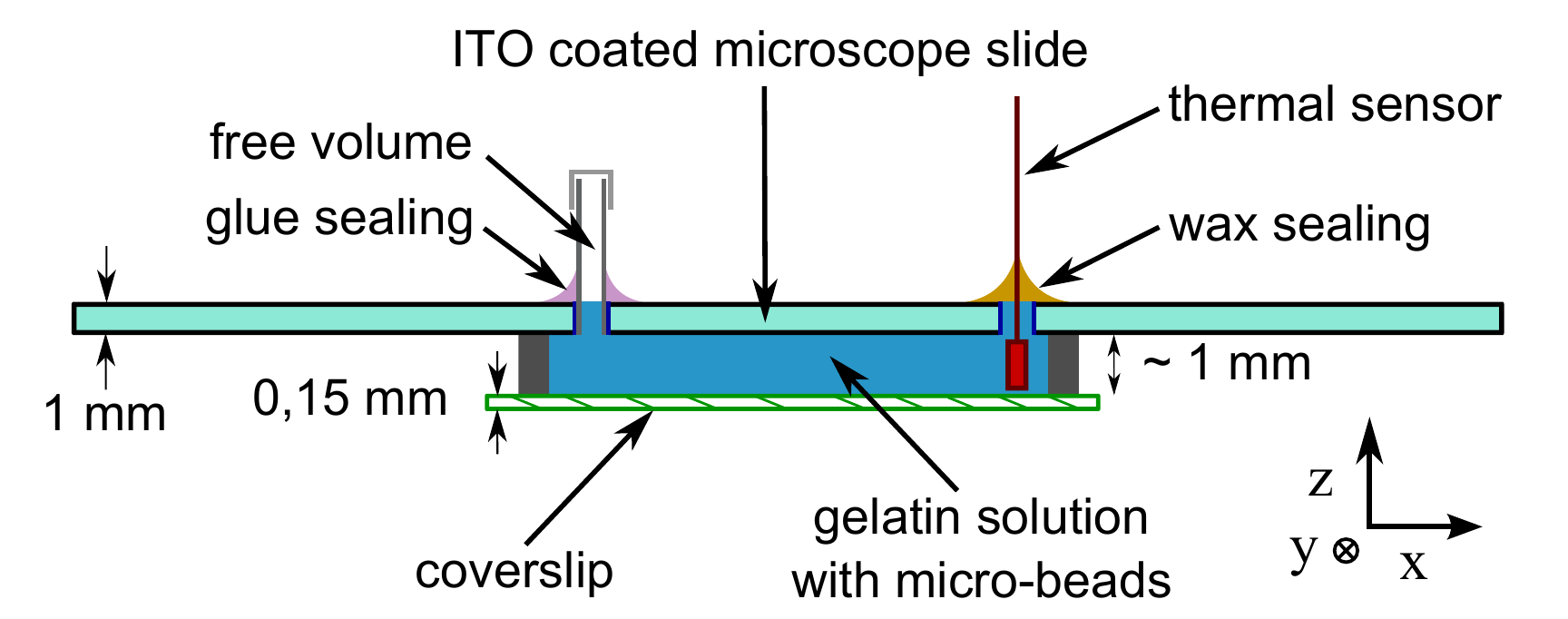}
\caption{Schematic representation of the cell used to trap particles in gelatin solution (view from the side). The microscope slide is ITO-coated, which enables us to heat the cell by sending an electrical current through the glass surface.} \label{gel:fig:cell}
\end{center}
\end{figure}

We use a disk-shaped glass cell, with an Indium Tin Oxyde (ITO) coated microscope slide, a free-volume to avoid problems if the volume of solution changes during gelation, and a \textit{Wavelength Electronics} TCS10K5 thermal sensor for temperature measurement (see figure \ref{gel:fig:cell}). To fill the cell, the gelatin solution with dispersed silica particles is taken from the refrigerator and heated at $\sim \SI{50}{\celsius}$ until it is in the sol phase.

\subsection{Optical trapping and controlled gelation}
\label{gel:section:optical_trapping}

\begin{figure}[ht!]
\begin{center}
\includegraphics[width=12cm]{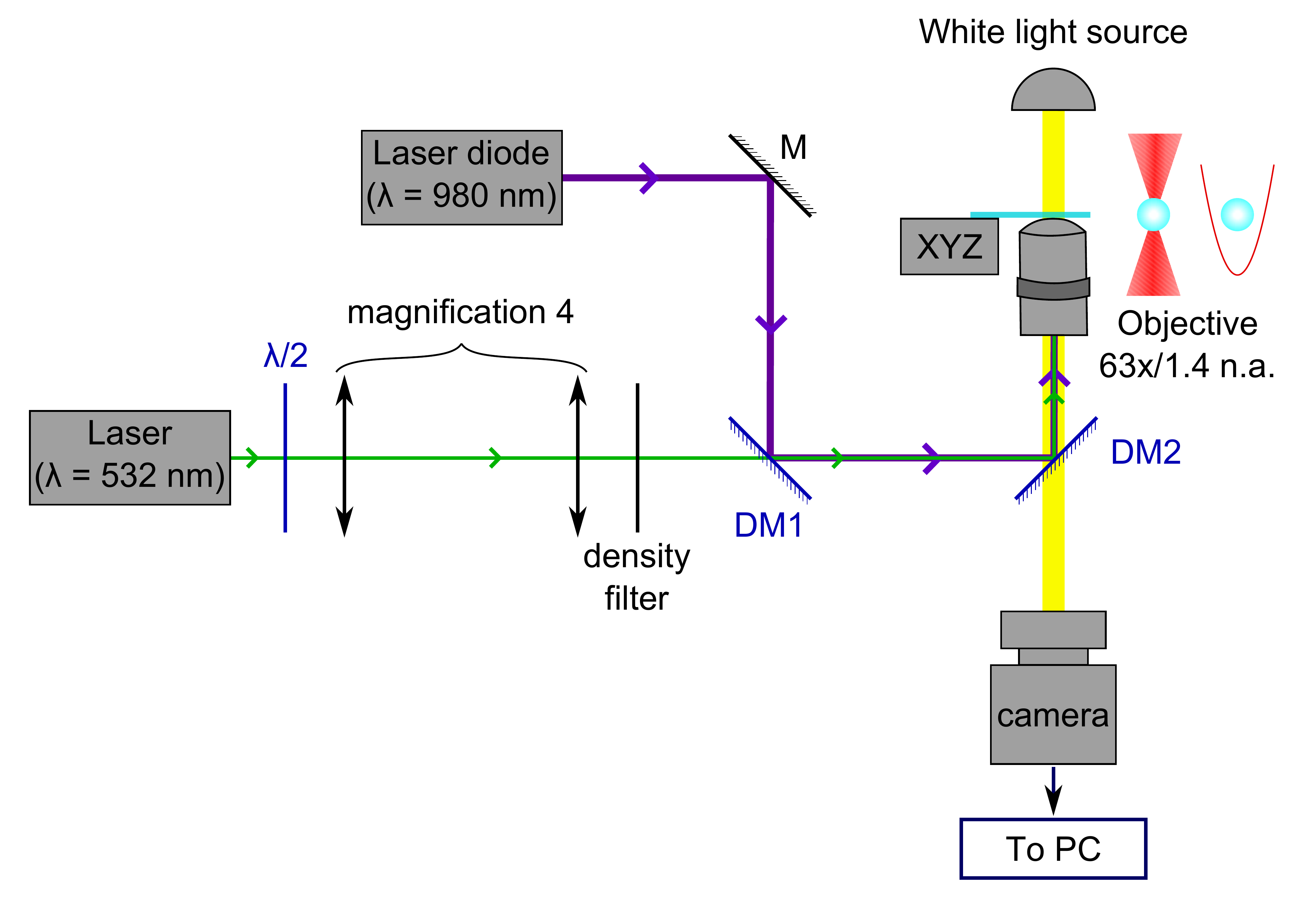}
\caption{Schematic representation of optical tweezers set-up used to trap two particles in gelatin solution with a green laser. An infrared laser diode is used to melt the gelatin. ``M'' is a mirror and ``DM'' are dichroic mirrors.} \label{gel:fig:setup_laser_vert}
\end{center}
\end{figure}

For different purposes we used two different alignments of an optical tweezers set-up:

The first set-up uses a laser beam ($\lambda = \SI{532}{\nano\meter}$) separated in two cross-polarized beams which enables us to have two traps with no interference between them. A laser diode ($\lambda = \SI{980}{\nano\meter}$) is aligned with the green laser and used to heat locally the sample. The tracking is done using a fast camera which is able to track two particles at \SI{600}{\hertz}. The acquisition rate is not very fast but by previously using a calibration target, we can directly convert the displacement of the particle from pixels to \si{\micro\meter}. See figure~\ref{gel:fig:setup_laser_vert}.

\begin{figure}[ht!]
\begin{center}
\includegraphics[width=12cm]{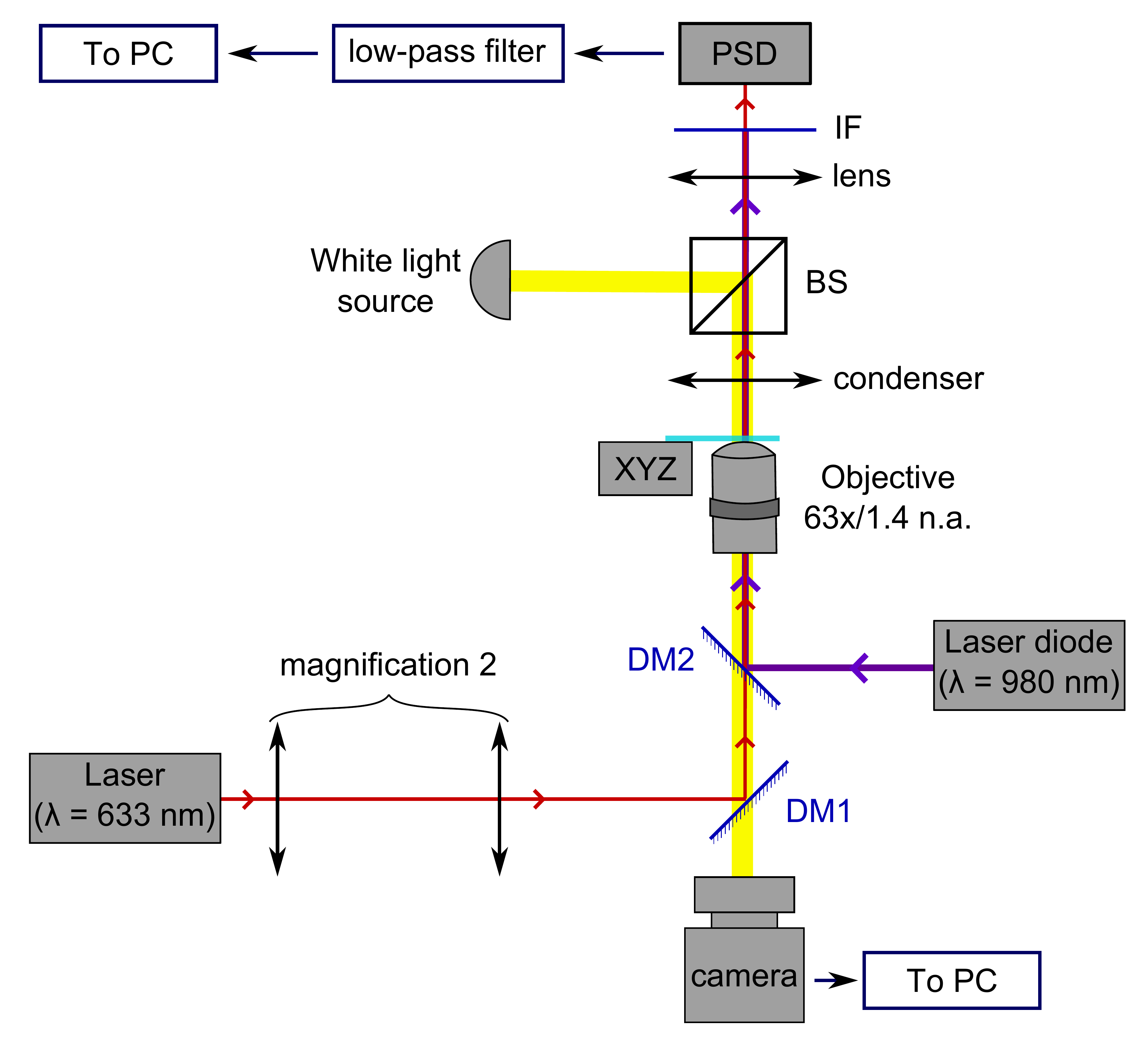}
\caption{Schematic representation of optical tweezers set-up to trap one particle in gelatin solution. An infrared laser diode is used to trap and to melt the gelatin. The deflection of a He-Ne laser induced by the trapped bead is measured by the position sensing diode (PSD). The white light source and camera are only used for direct visualization but not for the measurements. ``BS'' is a beam-splitter, ``IF'' and interferential filter to suppress the infrared beam and ``DM'' are dichroic mirrors.} \label{gel:fig:setup_laser_PSD}
\end{center}
\end{figure}

The second set-up uses a single laser diode ($\lambda = \SI{980}{\nano\meter}$) to trap the particle and to heat locally the sample. A He-Ne laser ($\lambda = \SI{632.8}{\nano\meter}$) is aligned with the laser diode and deflected by the particle. This deflection is measured using a position sensing diode which is able to track one particle at more than \SI{10}{\kilo\hertz}. Contrary to the first set-up, the position signal is in arbitrary unit and a supplementary calibration is needed for each measurement to convert the trajectory of the particle in physical units. See figure~\ref{gel:fig:setup_laser_PSD}.

For both set-ups, the microscope objective is an oil-immersion \textit{Leica} HCX PL. APO $\times 63$ with high numerical aperture $N.A. = 1.4$. The microscope objective is surrounded with a custom-made heating ring, made with a \textit{Minco} flexible resistor and a \textit{Wavelength Electronics} TCS10K5 thermal sensor for temperature measurement. A feedback control is managed by the temperature control module (TCM-39032) of a modular laser diode controller (ILX Lightwave LDC-3900). As already mentioned, another thermal sensor is inserted directly inside the cell (see figure~\ref{gel:fig:cell}) and another feedback control is done by mastering the current going through the ITO-coated microscope slide, with an \textit{Instec} MK1 Board and a PID software. These two temperature devices ensure that the temperature of both the microscope objective and the cell are well controlled. The precision achieved on the temperature control is about $\pm \SI{0.05}{\celsius}$.

To trap particles, the temperature of both the microscope objective and the cell are set to \SI{38}{\celsius} so that the gelatin is in the sol phase. Then, one  particle is found and trapped at a given distance from the bottom surface of the cell (typically $h = \SI{15}{\micro\meter}$). Then, the temperature controls are set to a value below the gel transition (typically $T_{\text{fb}} = \SI{27}{\celsius}$) and we let the sample gel for a few hours (typically between 6 and 10 hours). Since the gelatin shows a lot of hysteretic behaviour~\cite{djabourov1988gelationI,djabourov1988gelationII}, this gelation procedure appears to be important and the rheology of the gel can vary if the gelling time is very different (\textit{e.g.} a few days). Moreover, one must pay attention to regularly check the distance between the bead and the bottom surface of the cell, since the focal distance of the microscope objective always drifts slowly when its temperature is changed.

The refractive index of the liquid gelatin solution was measured $n_{\text{gel}} = 1.3415$, which is close to value in water $n_{\text{water}} = 1.3335$. It follows that the trapping stiffnesses in gelatin solution should be close to the ones in water with same experimental parameters.

\subsection{Local heating and fast quenching method}
\label{gel:section:fast_quench_method}

When the particle is trapped and the sample is properly gelled at a given controlled temperature $T_{\text{fb}} < T_{\text{gel}}$, the local quenches are done in a similar way than the one presented in~\cite{TheseRuben,RubenPRL2011,RubenEPL2012}:
The power of the \SI{980}{\nano\meter} laser diode is risen to a high value (typically\footnote{This is the power measured on the beam before the microscope objective, so the ``real'' power at the focal point should be smaller, due to the loss in the objective.} \SI{230}{\milli\watt}) during a given time (typically $\tau_{\text{melt}}=\SI{200}{\second}$). Because of the light absorption of the water molecules in the solution, the temperature of the gelatin around the particle (which is at the focal point of the microscope objective) increases by a small amount $\delta T$. Following the formula in reference~\cite{peterman2003laser}:
\begin{equation}
\delta T = \frac{P \alpha}{2 \pi K} \left[ \ln\left( \frac{2 \pi h}{\lambda} \right) - 1 \right]
\end{equation}
where $\alpha = \SI{50}{\per\meter}$ is the attenuation coefficient of water at \SI{27}{\celsius} for wavelength \SI{980}{\nano\meter}~\cite{Palmer1974}, and $K = \SI{0.61}{\watt\meter^{-1}\kelvin^{-1}}$ is the thermal conductivity of water. Here, we await: 
\begin{equation}
\delta T \simeq \SI{11}{\celsius}.
\end{equation}
This increase in temperature is only roughly estimated. Especially because we don't really know what is the absorption of the microscope objective for the near infrared, and because it is impossible to measure the temperature with a usual probe on this very small scale. But it is seen that the increase is strong enough to melt a small droplet of gelatin (radius $R_{\text{d}}\sim \SI{10}{\micro\meter}$) around the bead.
Then, the power is quickly decreased to a low value (in the case where the same laser diode is used to trap and heat) or to zero (in the case where another laser is used to trap the particle), and the sample is let gel for a given time (typically $\tau_{\text{rest}}=\SI{500}{\second}$). Since the thermal diffusivity of water is $\kappa = \SI{0.143e-6}{\square\meter\per\second}$ at \SI{25}{\celsius}, the time $\tau_{\kappa}$ needed to dissipate the heat from the droplet to the bulk is short:
\begin{equation}
\tau_{\kappa} \sim \frac{R_{\text{d}}^2}{\kappa} \sim \SI{2e-4}{\second}.
\end{equation}
Hence, the gelatin is believed to experience a fast quench at temperature $T_{\text{fb}} < T_{\text{gel}}$ and should start ageing\footnote{Actually, in the case where the same laser diode is used for trapping and melting the droplet, the temperature of the quench is a little bit above $T_{\text{fb}}$ because of the absorption of the laser. Since the power of the laser is low, this increase is less than \SI{1}{\celsius} and can easily be compensated by lowering $T_{\text{fb}}$ accordingly.}. After the resting time $\tau_{\text{rest}}$ at low temperature $T_{\text{fb}}$, the power of the laser diode is risen again, and another quench is done. Note that the exact duration of $\tau_{\text{rest}}$ was not considered as important, because it was believed that the melting ``resets'' the gelatin sample and that all the anomalous behavior occurs right after the quench.

The position of the particle trapped in the center of the melted droplet is continuously measured during a succession of several melting and aging. For each measurement the quenching is repeated a few hundred times in order to perform proper ensemble averages.

\begin{figure}[ht!]
\begin{center}
\includegraphics[width=10cm]{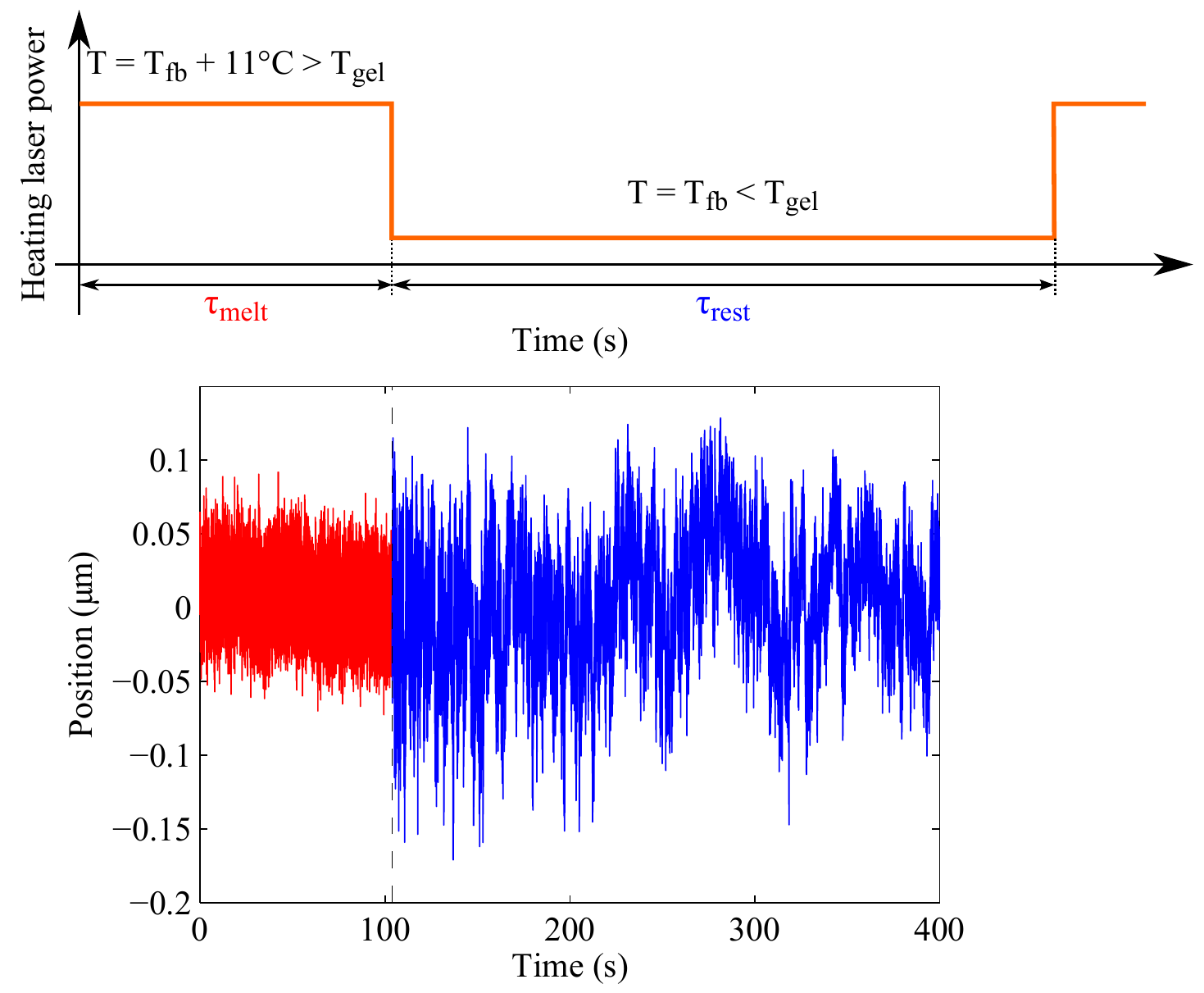}
\caption{Trajectory of one particle trapped in a gelatin sample kept at $T_{\text{fb}} = \SI{27.5}{\celsius}$, around the quench. On the red part of the trajectory, the intensity of the laser is high and the gelatin droplet is liquid. On the blue part, the intensity of the laser is low and the gelation is occuring.} \label{gel:fig:exemple_fonte}
\end{center}
\end{figure}

An example of trajectory obtained with the second set-up is presented in figure~\ref{gel:fig:exemple_fonte}. When the intensity of the laser is high, the gelatin droplet is in the ``sol'' phase and the particle fluctuates  in an optical trap with a high stiffness. When the intensity of the laser is low, the gelation is occurring and the stiffness of the trap is low (which is the reason why the position fluctuations are bigger).

\newpage

\section{Results}

In this section we present some results showing that there are no anomalous fluctuations occurring in the ageing of gelatin solution right after a fast quench. We also discuss why this effect that was previously observed is likely to actually be an artefact due to data analysis.

\subsection{Time evolution of bulk properties and hysteresis}

We first performed  preliminary measurement of gelatin gelation in bulk (\textit{i.e.} without the local heating method). We prepare a cell with gelatin solution as described in \ref{gel:section:sample_preparation}. We set the temperature controls of both the objective and of the cell at \SI{37}{\celsius} for \SI{30}{\minute} to melt the gelatin. We trap one particle and we switch both temperature controls to a given temperature $T_{\text{fb}}$. We wait a few tens of minutes (typically $\sim \SI{30}{\minute}$) and we set the distance between the particle and the bottom of the cell when there is no more drift due to thermal expansion of the microscope objective. After that, we measure the position of the trapped bead for a long time (\textit{e.g.} \SI{8}{\hour}) at \SI{400}{\hertz} to see the bulk gelation of the sample.

For all these measurements, the distance between the bead and the surface is $h=\SI{15}{\micro\meter}$. The liquid gelatin solution at \SI{37}{\celsius} is a Newtonian fluid, and the stiffness $k$ of the trap can be computed directly from the variance of the position $x$ of the bead:
\begin{equation}
\sigma_{x}^2 = \left\langle x^2\right\rangle = \frac{k_{\text{B}}T}{k}
\end{equation}
For all these measurement the stiffness of the trap was $k = \SI{0.46(1)}{\stiffness}$.

Since the $T_{\text{gel}}$ is expected to be around \SI{29}{\celsius}, we varied $T_{\text{fb}}$ from \SI{31}{\celsius} to \SI{27.5}{\celsius}. It was found that above \SI{28.5}{\celsius}, the gelation does not occur on the time of the experiment and the solution stays liquid, even if its viscosity increases continuously. Below \SI{28.3}{\celsius}, the gelation occurs before the end of the experiment. It was estimated that the bulk gelation of the cell volume takes $\sim \SI{260}{\minute}$ at \SI{28.3}{\celsius} and $\sim \SI{120}{\minute}$ at \SI{27.5}{\celsius}.

\begin{figure}[ht!]
\begin{center}
\includegraphics[width=9cm]{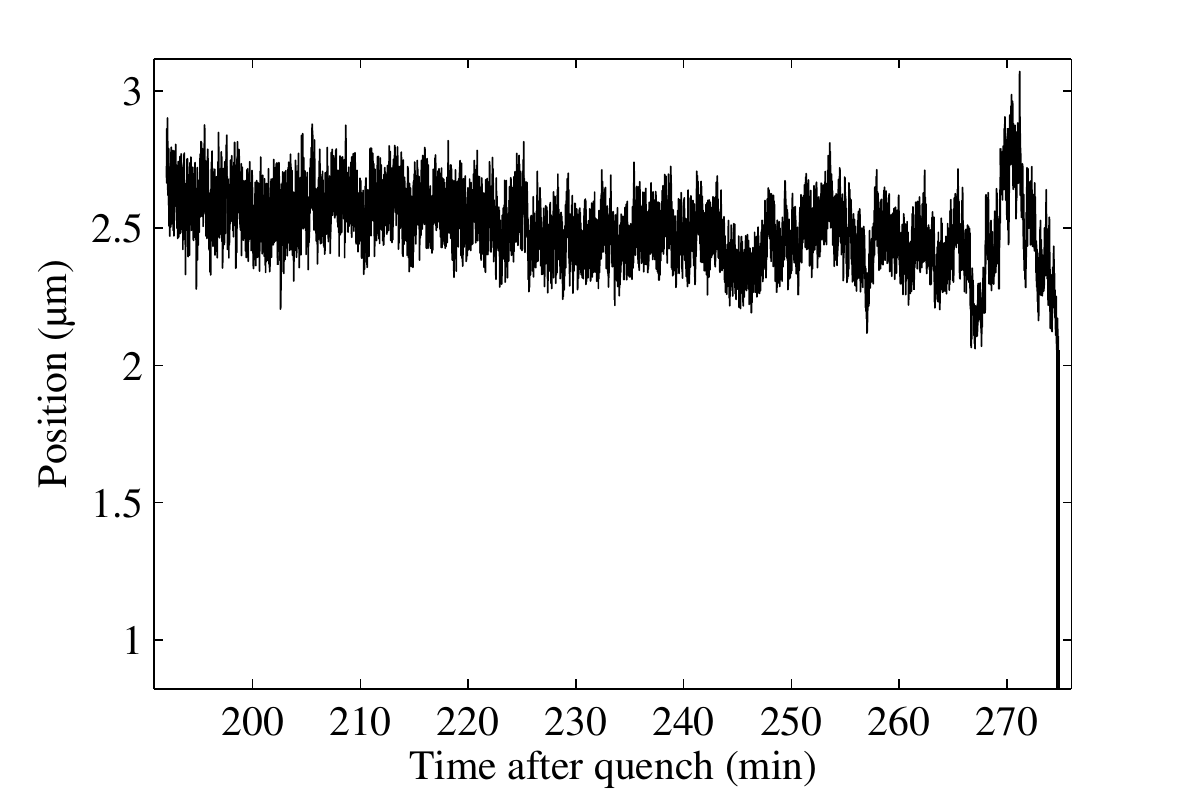}
\caption{Evolution of the position of one trapped particle, in gelatin solution kept at \SI{28.3}{\celsius} (after being melt at \SI{37}{\celsius} for \SI{20}{\minute}). At the end of the trajectory, gelation occurs and the particle is moved away from the optical trap.} \label{gel:fig:gelation_trajectory}
\end{center}
\end{figure}

Estimating the state (``sol'' or ``gel'') of the gelatin solution, is not trivial, since the fluid can be really viscous without being completely gelled. Qualitatively, the trajectory of the trapped bead starts to be heckled, and the bead sometimes escapes the trapping (see figure~\ref{gel:fig:gelation_trajectory}). An \textit{a posteriori} test consists in switching off the laser (resulting in switching off the trapping) and letting the sample at $T_{\text{fb}}$ for a few more hours (typically over night) to see if the particle slowly fall to the bottom of the cell. If the particle does not fall, the gelatin solution is considered to be fully gelled.

To estimate the evolution of the viscosity during the gelation process, we used passive micro-rheology techniques~\cite{Schmidt2008}. The trajectories were divided in portions of $\sim \SI{1}{\hour}$, and the Power Spectral Density (PSD) was computed for each portion. A long trajectory is required because we need low frequencies to correctly estimate the PSD. We explicitly assume that the bulk aging is slow enough for not perturbing too much the estimation of the PSD when a long trajectory is taken. Or at least, that taking a long time-window will only smooth the rheology result. 

\begin{figure}[ht!]
        \centering
        \begin{subfigure}[b]{0.5\textwidth}
                \includegraphics[width=\textwidth]{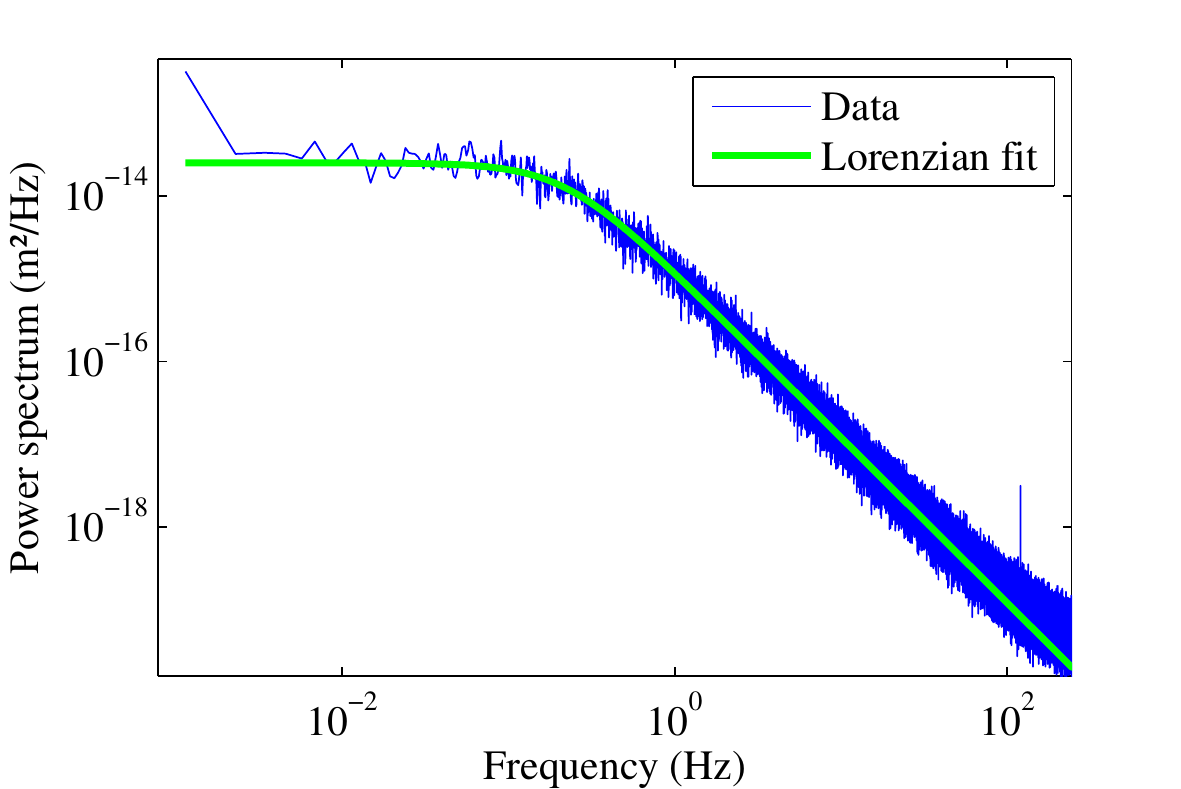}
                \caption{\SI{70}{\minute} after the temperature change.}
                \label{gel:fig:spectrum_70min}
        \end{subfigure}%
        \begin{subfigure}[b]{0.5\textwidth}
                \includegraphics[width=\textwidth]{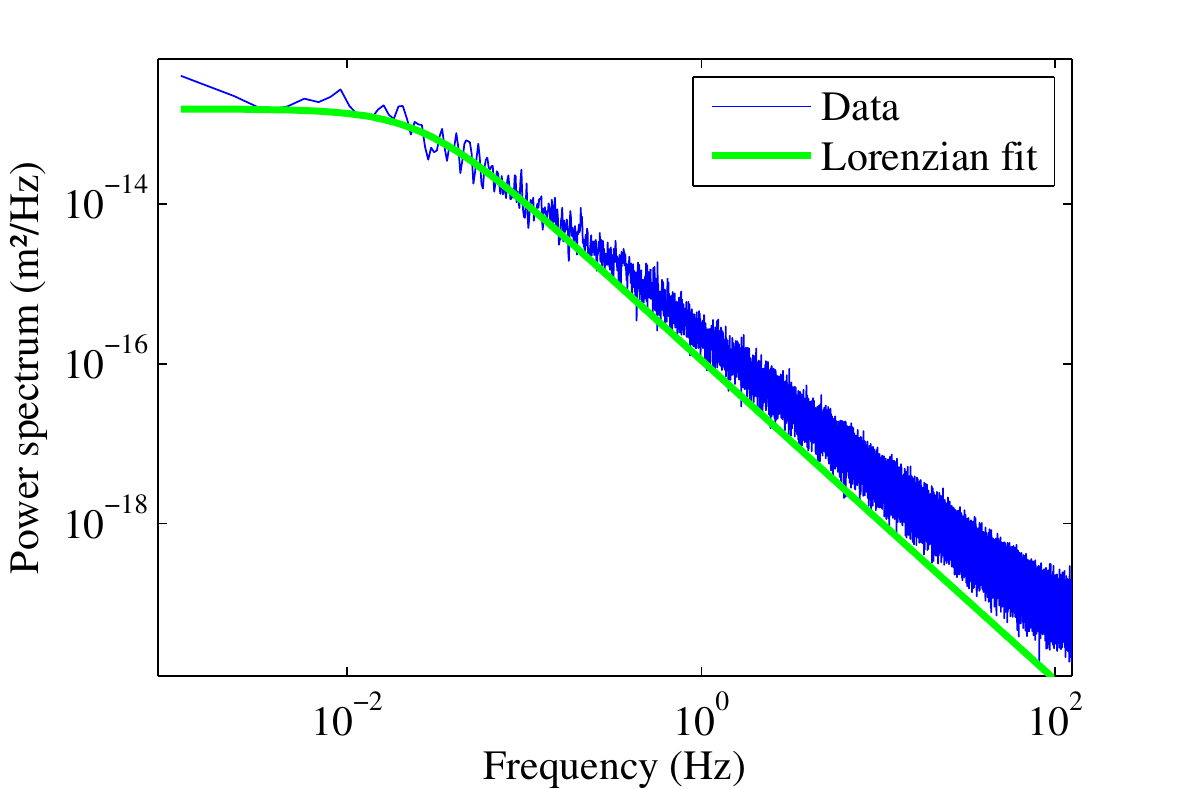}
                \caption{\SI{320}{\minute} after the temperature change.}
                \label{gel:fig:spectrum_320min}
        \end{subfigure}
        \caption{Power Spectral Density of the position of one particle trapped in gelatin solution kept at \SI{28.5}{\celsius} (after being melt at \SI{37}{\celsius} for \SI{30}{\minute}). The PSD is estimated over a time-window of \SI{1}{\hour}. Shortly after switching the temperature the gelatin solution is still a Newtonian fluid and the PSD is Lorentzian. After some time, visco-elastic effects appear and the PSD is no longer Lorentzian.}\label{gel:fig:spectrums_bulk}
\end{figure}

As seen in figure~\ref{gel:fig:spectrum_70min}, shortly after the decrease of temperature, the PSD is still Lorentzian, as awaited for a particle trapped in a Newtonian fluid at equilibrium~\cite{BergSorensenRSI2004}. The viscous drag coefficient $\gamma = 6\pi R \eta$ (with $\eta$ the dynamical viscosity of the solution) can be estimated from the value of the cut-off frequency $f_c=k/(2\pi\gamma)$. Here we find: $\eta = \SI{21(1)e-3}{\pascal\second}$. As the gelation occurs, the PSD is less and less Lorentzian (see figure~\ref{gel:fig:spectrum_320min}), which is the sign that the gelatin solution starts to behave as a viscoelastic fluid~\cite{Sasso2009}.

\begin{figure}[ht!]
\begin{center}
\includegraphics[width=9cm]{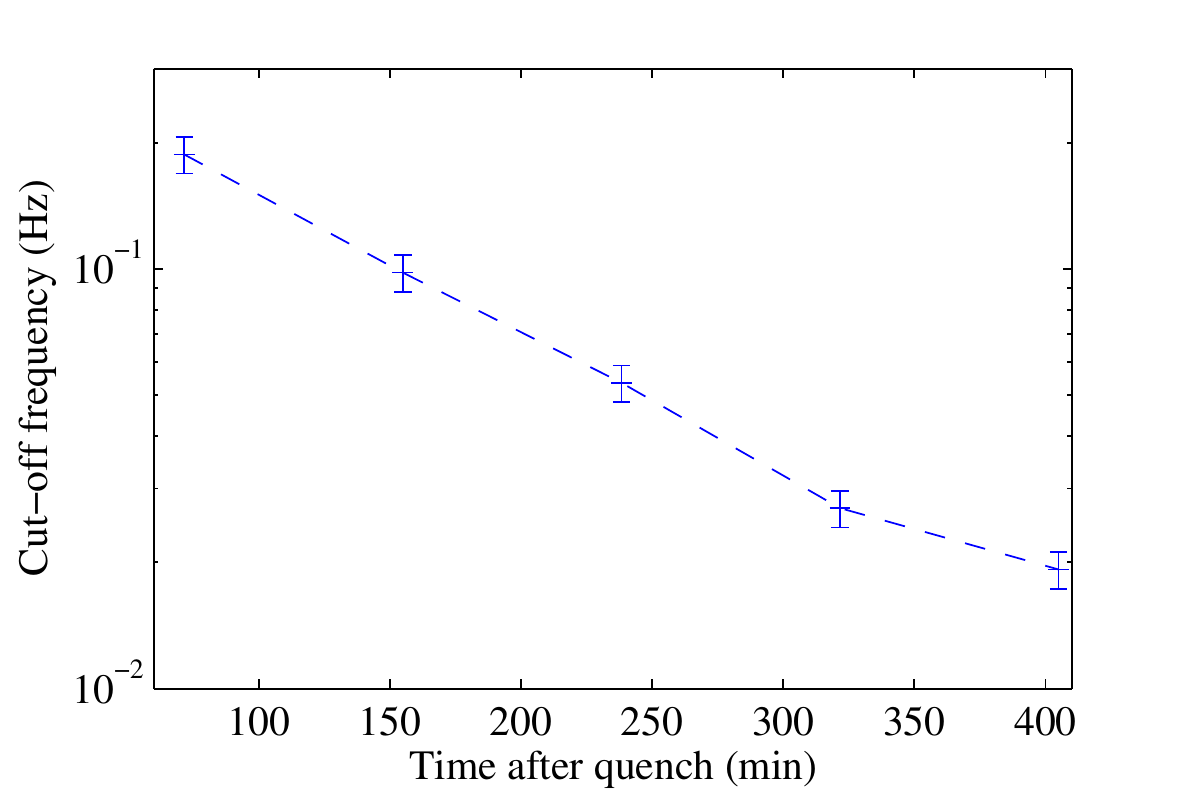}
\caption{Evolution of the fitted cut-off frequency $f_c$, in gelatin solution kept at \SI{28.5}{\celsius} (after being melt at \SI{37}{\celsius} for \SI{30}{\minute}).} \label{gel:fig:fc_bulk}
\end{center}
\end{figure}

We plot in figure~\ref{gel:fig:fc_bulk} the evolution of the fitted cut-off frequency at different time after the gelatin solution was set at $T_{\text{fb}}=\SI{28.5}{\celsius}$. Even if the spectrum is no longer Lorentzian near the end of the measurement, it seems that the cut-off frequency decreases exponentially. Therefore, the apparent viscosity increase is exponential.

From these preliminary measurements, we estimate that the $T_{\text{gel}}$ is about \SI{28.3}{\celsius} for our gelatin solution at \SI{5}{\weight}. We chose to work with $T_{\text{fb}} < \SI{28.3}{\celsius}$ for all the following quenching experiments. As mentioned earlier, gelatin solutions have big hysteretic behaviour~\cite{djabourov1988gelationI,djabourov1988gelationII}. It follows that the viscoelastic properties of the solution in an important temperature range around $T_{\text{gel}}$ cannot be known independently of the sample's history.

\begin{figure}[ht!]
        \centering
        \begin{subfigure}[t]{0.5\textwidth}
                \includegraphics[width=\textwidth]{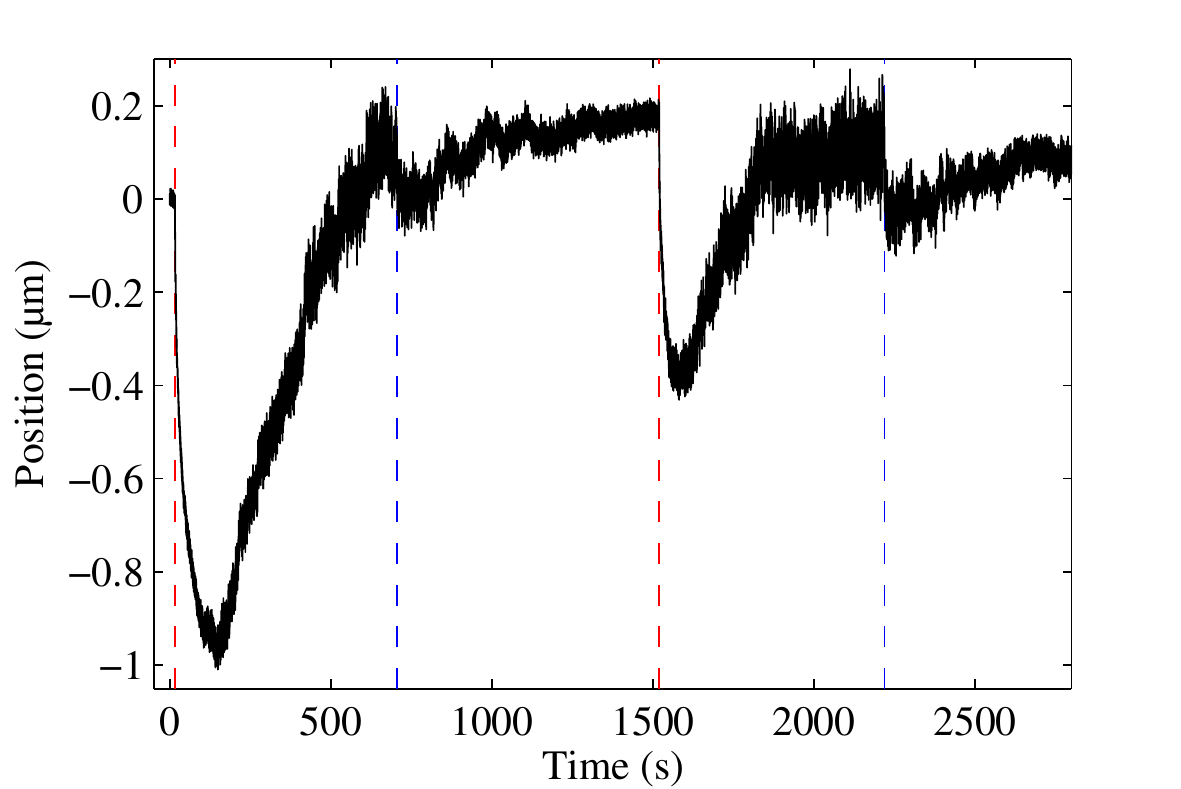}
                \caption{First cycles.}
                \label{gel:fig:exemple_mauvais_fonte_1}
        \end{subfigure}%
        \begin{subfigure}[t]{0.5\textwidth}
                \includegraphics[width=\textwidth]{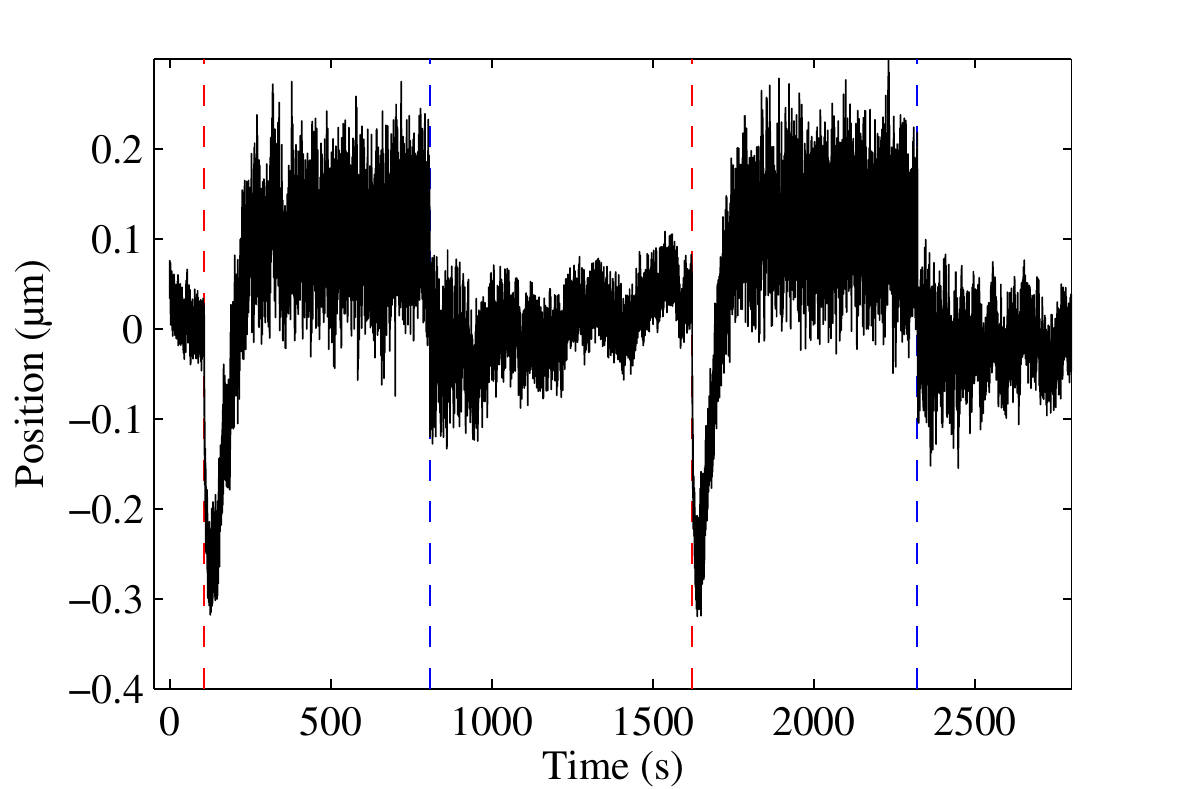}
                \caption{\SI{3}{\hour} after starting the cycles.}
                \label{gel:fig:exemple_mauvais_fonte_2}
        \end{subfigure}
        \caption{Examples of trajectories for melting/regelling cycles with a gelatine sample gelled at $T_{\text{room}} = \SI{24}{\celsius}$ for a long time. The red dashed-lines indicate when the heating laser is switched ON, and the blue ones when it is switched OFF. At the beginning, the melting is more difficult to reach, after a given time, the cycles look ``reproducible''.}\label{gel:fig:exemple_mauvais_fonte}
\end{figure}

Another consequence of the hysteretic behaviour is that the first bulk gelation of the sample must be done in a controlled and reproducible manner. If the sample is let gel for a too long time (generally more than one day), or at a too low temperature ($\lesssim \SI{22}{\celsius}$), the first melting/regelling cycles used for the quenching experiment will be different from the following ones (where a ``reproducible'' state is reached). Especially, in this case the first melting is more difficult to reach. Examples of trajectories are shown in figure~\ref{gel:fig:exemple_mauvais_fonte}. One can clearly see some position drifts occurring when the temperature is increased, before reaching a ``sol'' state where the particle fluctuates in the optical trap. Note that, contrary to figure~\ref{gel:fig:exemple_fonte}, we use here the first set-up where the trapping laser is different from the heating laser (and the focal point of the two laser are not aligned). Hence, there is nearly no change of the trap stiffness when the intensity of the heating laser is changed. The change of fluctuations amplitude when the heating laser is switched OFF is mostly due to the rapidly increase of gelatin viscosity when gelation occurs. 

\subsection{Difference between ensemble variance and temporal variance in the presence of a drift}

We now consider quenching experiment as described in~\ref{gel:section:fast_quench_method}. We obtain several temporal trajectories of the particles positions for a given quenching temperature $T_{\text{fb}} < T_{\text{gel}}$. The important point is to estimate correctly the statistical properties from this set of data. In particular, we are interested in the variance of the position, which has been seen to have an anomalous increase right after the quench~\cite{TheseRuben,RubenPRL2011,RubenEPL2012}.

\begin{figure}[ht!]
\begin{center}
\includegraphics[width=8cm]{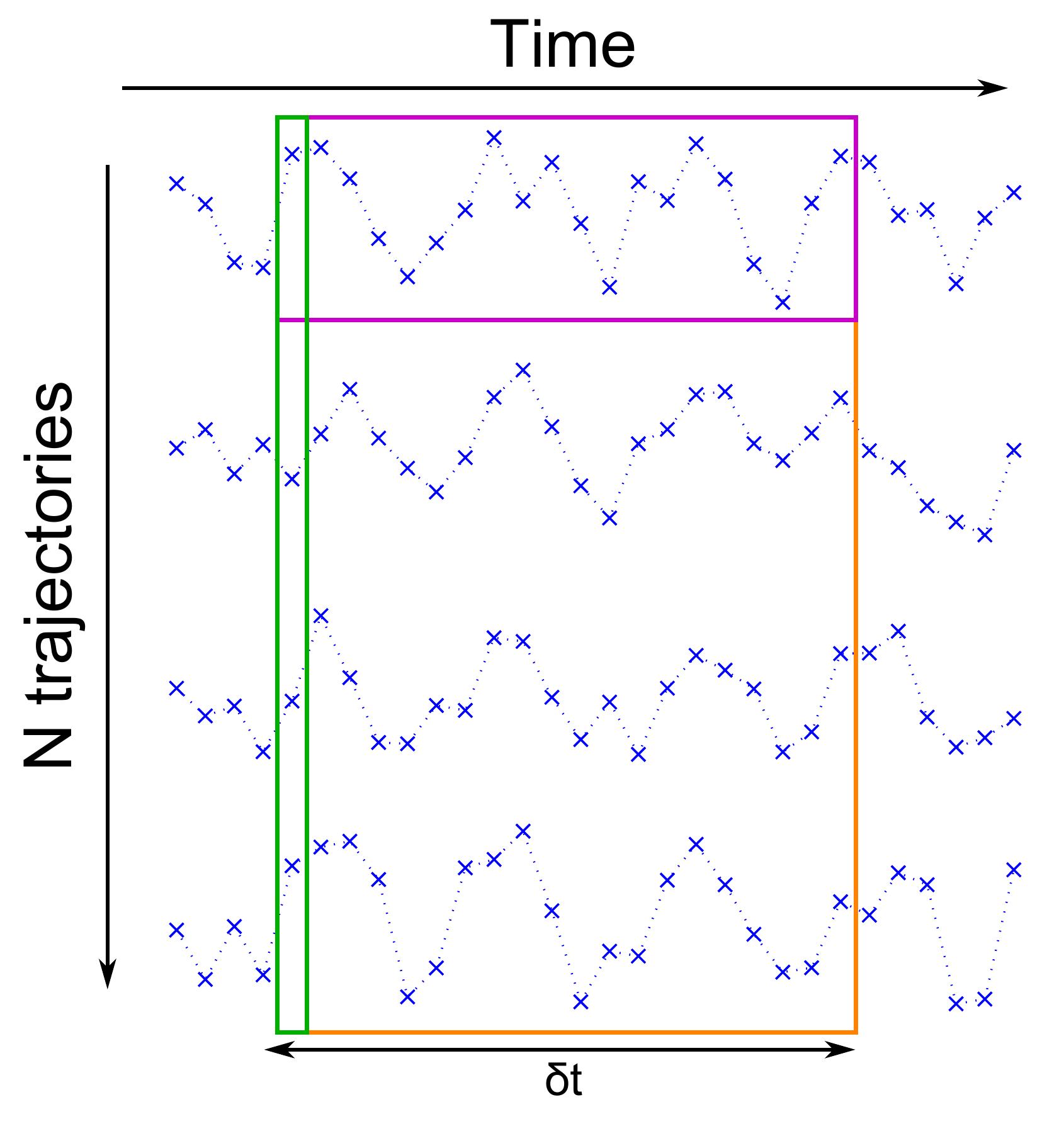}
\caption{Schematic representation of the different ways to estimate the variance for a set of N trajectories with a time-window $\delta t$. The temporal variance $\sigma_{\text{time}}^2$ is computed by estimating the variance of the points in the \textcolor[RGB]{200,0,200}{fuschia box}, and then averaging over the trajectories. The ensemble variance $\sigma_{\text{ensemble}}^2$ is computed by estimating the variance of the points in the \textcolor[RGB]{0,175,0}{green box}, and then averaging over the time-window $\delta t$. The boxed variance $\sigma_{\text{box}}^2$ is computed directly by estimating the variance of all the points in the \textcolor[RGB]{255,130,0}{orange box}.} \label{gel:fig:definition_variances}
\end{center}
\end{figure}

The correct ensemble variance should be estimated instantaneously at a given time $t$, by considering the $N$ different trajectories at this time $t$. If one wants to increase the statistics by taking a small time-window $\delta t$, there are at least 3 ways to compute the variance from the set of trajectories. These different ways are schematically represented on figure~\ref{gel:fig:definition_variances}. We call $x_{i}(t)$ the position of the particle for the $i^{\text{th}}$ quench at the time $t$:
\begin{itemize}
\item The temporal variance $\sigma_{\text{time}}^2$ is obtained by estimating the variance over the time $\delta t$ for each quench, and then averaging over the N quenches:
\begin{equation}
\sigma_{\text{time}}^2 (t)= \frac{1}{N} \sum_{i=1}^{N} \left[ \frac{1}{\delta t} \int_{t}^{t+\delta t} \left( x_{i}(t^{\prime})-\bar{x}_{i} (t) \right)^2 \, \mathrm dt^{\prime} \right]
\label{gel:eq:temporal_variance}
\end{equation}
where $\bar{x}_{i} (t) = \frac{1}{\delta t} \int_{t}^{t+\delta t} x_{i}(t^{\prime}) \, \mathrm dt^{\prime}$ is the temporal mean of $x$ for the $i^{\text{th}}$ quench, between $t$ and $t+\delta t$.
\item The ensemble variance $\sigma_{\text{ensemble}}^2$ is obtained by estimating the variance over the N quenches at a time $t$ and then averaging over the time-window $\delta t$:
\begin{equation}
\sigma_{\text{ensemble}}^2 (t)= \frac{1}{\delta t} \int_{t}^{t+\delta t} \left[ \frac{1}{N-1} \sum_{i=1}^{N} \left( x_{i}(t^{\prime})-\langle x (t^{\prime}) \rangle \right)^2 \right] \, \mathrm dt^{\prime}
\label{gel:eq:ensemble_variance}
\end{equation}
where $\langle x (t^{\prime}) \rangle = \frac{1}{N} \sum_{i=1}^{N} x_{i} (t^{\prime})$ is the ensemble mean of the N trajectories $x_{i}(t^{\prime})$ at time $t^{\prime}$.
\item The boxed variance $\sigma_{\text{box}}^2$ is obtained by taking the N segments of trajectory from $x_i(t)$ to $x_i(t+\delta t)$, and then estimating the variance of the whole set of data:
\begin{equation}
\sigma_{\text{box}}^2 (t) =  \frac{1}{N\delta t} \sum_{i=1}^{N} \int_{t}^{t+\delta t} \left( x_{i} (t^{\prime}) - \boxed{x} (t) \right) ^2   \, \mathrm dt^{\prime}
\end{equation}
where $\boxed{x} (t) = \frac{1}{N\delta t} \sum_{i=1}^{N} \int_{t}^{t+\delta t} x_{i} (t^{\prime}) \, \mathrm dt^{\prime}$ is the mean computed on the set of data made of the N segments from $x_i(t)$ to $x_i(t+\delta t)$.
It is the variance used in references~\cite{TheseRuben,RubenPRL2011,RubenEPL2012}. 
\end{itemize}
\textit{Nota Bene}: Here to clearly distinguish the role of the time and the ensemble averages we have considered the time as a continuous variable and the number of trajectories as discrete. But experimentally the time is of course also a discrete variable, since we take measurements with a finite sampling frequency.

If the system is at equilibrium and $\delta t$ is big enough to correctly take account of the low-frequency of the motion, all these values should be equal to the equipartition value $k_{\text{B}}T/k$, with $k_{\text{B}}$ the Boltzmann constant, $T$ the temperature and $k$ the trap's stiffness.

Unfortunately, when the system is non-stationary (which is the case for an ageing system), these 3 definitions of the variance are not equivalent. Especially, if there's a slow drift existing on each trajectory, the estimations that average over time (\textit{i.e.} temporal and boxed variances) are likely to show a strong artefact.

\begin{figure}[ht!]
\begin{center}
\includegraphics[width=9cm]{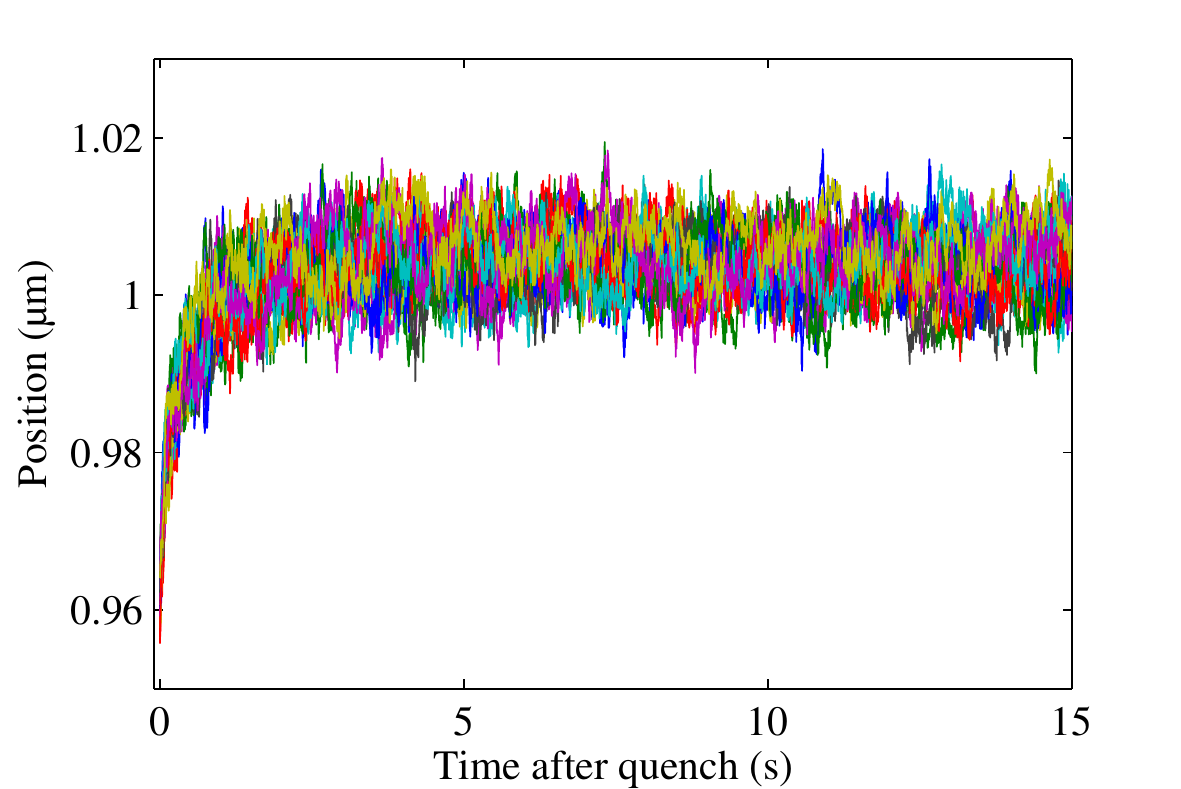}
\caption{20 first trajectories for a quench at $T_{\text{fb}} = \SI{28}{\celsius}$, sampled at \SI{400}{\hertz}. A slow drift of $\sim \SI{40}{\nano\meter}$ is clearly visible during the first $\sim \SI{1}{\second}$. After that, the position only oscillates randomly around a mean value.} \label{gel:fig:20_quenches}
\end{center}
\end{figure}

To illustrate this effect, we have taken a set of 178 quenches done with the first set-up described in~\ref{gel:section:optical_trapping} at \SI{28}{\celsius}, sampled at \SI{400}{\hertz}. The parameters were: melting time $\tau_{\text{melt}} = \SI{250}{\second}$, melting intensity $I_{\text{melt}} = \SI{235}{\watt}$, resting time $\tau_{\text{rest}} = \SI{305}{\second}$ and trap stiffness\footnote{The trap stiffness is measured when the gelatin sample is completely melt and kept at constant temperature $T=\SI{37}{\celsius}$, before the first bulk gelation.} $k = \SI{3.7}{\stiffness}$.
One can clearly see on the trajectories that there is a small drift of $\sim \SI{40}{\nano\meter}$ which occurs right after the quench (see figure~\ref{gel:fig:20_quenches}). Such a drift is often seen for this kind of measurement. We interpret it as a slow relaxation of the gel network, which occurs on a time much smaller than the gelation, but much greater than the heat dissipation. In other words, when the gelation occurs, the particle is trapped in the gel network at a given position. And even if we melt a small droplet, the gelatin network will somehow ``remind'' this position and pull the particle back to its place when it regels. Here the drift is very visible because the position of the trapping laser is not the same as the position of the locally heating laser. Thus the position where the particle was during the first bulk gelation is not the position where the particle is attracted to when the gelatin is melted. But even when there is only one laser used for both trapping and heating, this drift can occur. It is indeed impossible to verify that the position where the particle is when the sample gelled is exactly the position of the laser, and a drift of only a few \si{\nano\meter} can be visible. This kind of drift can be avoided by having a more powerful heating laser to completely melt the gelatin on a larger area, as in~\cite{Ruben2013}.

We then have three characteristic times :
\begin{itemize}
\item $\tau_{\text{gel}}$ the time needed for the gelatin solution to regel completely. It goes from a few hundreds to more than $\SI{1000}{\second}$ depending on the quench temperature $T_{\text{fb}}$. 
\item $\tau_{\text{dynamics}}$ the typical time of the particle motion, which is directly $\frac{1}{f_{c}}$ and evolves from $\sim 5$ to $\sim \SI{100}{\second}$ during the gelation process.
\item $\tau_{\text{drift}}$ the time where the drift is visible, which is typically \SI{1}{\second} for our experiment.
\end{itemize}

\begin{figure}[ht!]
        \centering
        \begin{subfigure}[b]{0.5\textwidth}
                \includegraphics[width=\textwidth]{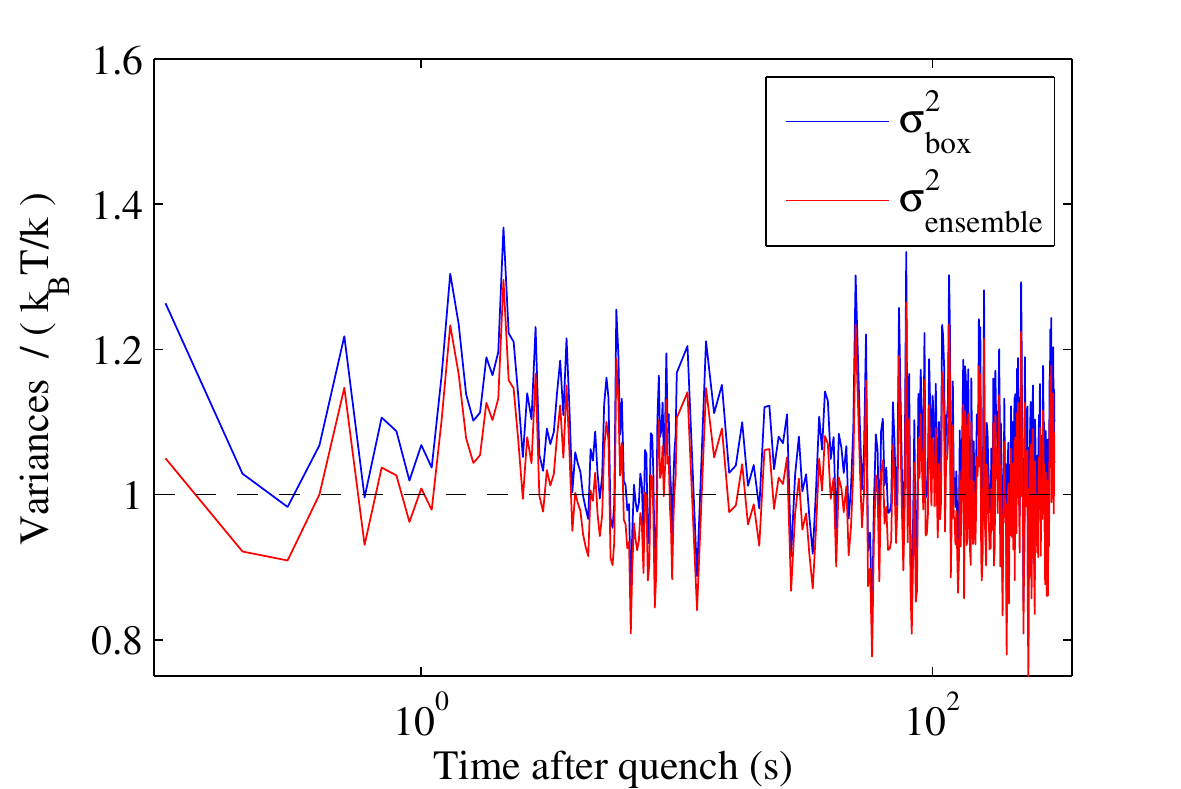}
                \caption{Ensemble and boxed variances.}
                \label{gel:fig:var_box_et_ens_01s}
        \end{subfigure}%
        \begin{subfigure}[b]{0.5\textwidth}
                \includegraphics[width=\textwidth]{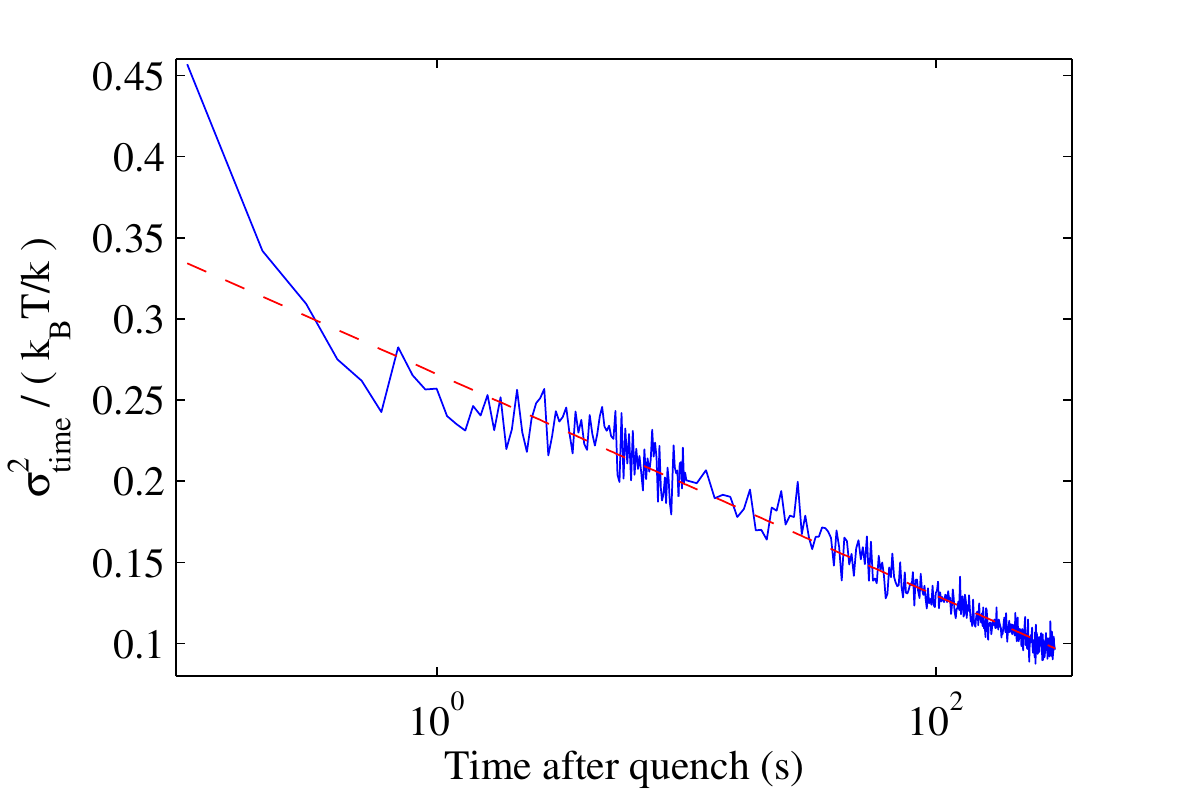}
                \caption{Temporal variance.}
                \label{gel:fig:var_temp_01s}
        \end{subfigure}
        \caption{Different variances computed for $\delta t = \SI{0.1}{\second}$ and normalised by the equilibrium value $k_{\text{B}}T/k$. The ensemble and boxed values are nearly equal and seem to be close to the equilibrium value at any time after the quench. Whereas the temporal value is clearly below the equilibrium value and decreases logarithmically with time after the quench.}\label{gel:fig:different_variances_01s}
\end{figure}

If we take a $\delta t$ sufficiently small compared to $\tau_{\text{drift}}$, the boxed and ensemble variances will give more or less the same result. Whereas, since $\tau_{\text{drift}} < \tau_{\text{dynamics}}$, it is clear that the temporal variance will dramatically underestimate the variance due to the lack of low frequencies signal. Indeed, the temporal variance would require a $\delta t$ of the order of magnitude of $\tau_{\text{dynamics}}$ for a correct estimation, which cannot be used because of the drift and the aging. Data are shown on figure~\ref{gel:fig:different_variances_01s} for $\delta t =\SI{0.1}{\second}$.

\begin{figure}[ht!]
\begin{center}
\includegraphics[width=9cm]{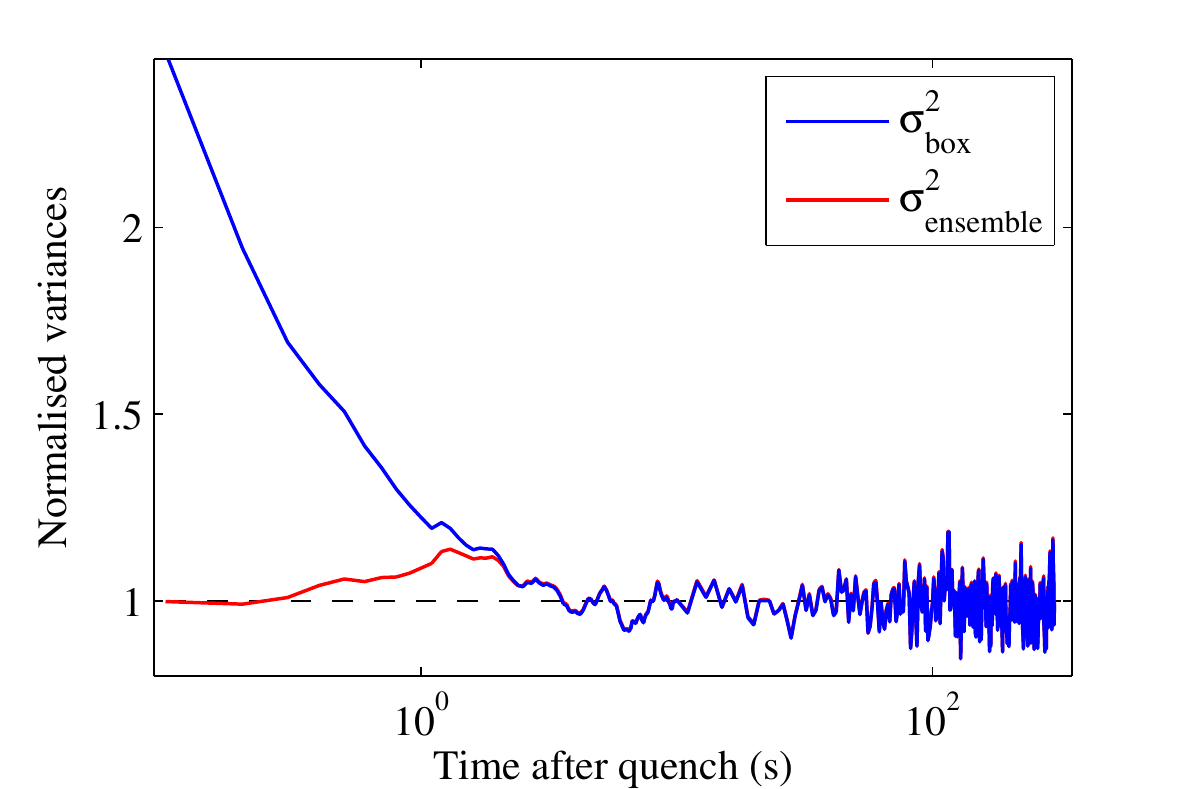}
\caption{Ensemble and boxed variances computed for a $\delta t = \SI{1}{\second}$ and normalised by the equilibrium value $k_{\text{B}}T/k$. The boxed variance clearly shows an anomalous increase at small times after the quench, which is the effect reported in previous works.} \label{gel:fig:var_box_et_temp_1s}
\end{center}
\end{figure}

Now, if the chosen $\delta t$ is too big compared to the characteristic time of the drift, the boxed variances will start to show an anomalous increase. Data are shown on figure~\ref{gel:fig:var_box_et_temp_1s} for $\delta t =\SI{1}{\second}$. This increase is not a real non-equilibrium effect due to the sol-gel transition, but only an artefact due to data analysis in presence of a slow drift. However, this slow drift is due to the fact that the sample is a gelatin solution, where an elastic network is created in the ``gel'' phase.

It is nevertheless interesting to see that the correct ensemble variance seems to satisfy the equilibrium equipartition relation at any time after the fast quench, even though there is a clear evolution of the visco-elastic properties with time, and even in the presence of a slow drift at the beginning of the quench: 
\begin{equation}
\forall t: \sigma_{\text{ensemble}}^2 (t) = \frac{k_{\text{B}}T}{k}.
\label{gel:eq:constVar}
\end{equation}
Similar results were seen for different quenches temperatures from \SI{28}{\celsius} to \SI{26}{\celsius}.

In~\cite{TheseRuben,RubenPRL2011,RubenEPL2012} it is stated that for longer times after the quench, the variance should decrease because of the elasticity of the gelatin network (as seen figure~\ref{gel:fig:Ruben_1}). This result is not clear. Of course, the particle dynamics becomes arrested after gelation, and at some point the amplitude of its positional fluctuations must decrease in time. At the gel point, the storage modulus and the yield stress of the gelatin sample becomes non-zero, and even for a freely suspended particle the dynamics becomes subdiffusive~\cite{Larsen2008}. Therefore it is clear that the temporal variance calculated for each trajectory should go to zero at long time. However, the ensemble variance calculated instantaneously over several trajectories should not go to zero. Otherwise it would mean that the position where the particle stops to move is the same for every trajectory. Thus, the fact that the variance decreases for long time in previous works is a sign that the ensemble analysis is mixed with some temporal analysis.

It was verified experimentally that for $\tau_{\text{rest}}$ up to \SI{900}{\second} the correct ensemble variance remains constant. 
Afterwards, it is not clear that equation~\ref{gel:eq:constVar} is verified, but we do not observe a shrinkage of the distribution of positions.


\subsection{Correct Estimation of Position Distribution Function}

We now want to study not only the variance of the position fluctuations but also their complete Probability Distribution Function (PDF). In order to minimize the risk of slow drifts and to increase the sample frequency, we did new measurements with the second experimental set-up described in~\ref{gel:section:optical_trapping}. With this set-up, the trapping and heating laser are the same, and the sample frequency can go up to \SI{10}{\kilo\hertz}. However, the calibration of the measured deflection from \si{\volt} to \si{\micro\meter} requires a supplementary assumption (for example, that the Fluctuation-Dissipation Theorem is verified, as it is done in \cite{BergSorensenRSI2010}). We will start by doing no assumption and plot the results only in arbitrary units.

One must pay attention at which ``position fluctuation'' is considered, as people often look at the distribution of $\delta x = x - \langle x \rangle$. The $\langle x \rangle$ is the mean of the position $x$, which can be defined in several ways when the system is not a classic stationary ergodic system. Especially, when one considers a small time-window $\delta t$, the correct mean should be the ensemble average $\langle x (t) \rangle$ estimated for each time $t$ (as defined in equation~\ref{gel:eq:ensemble_variance}). But if one takes instead the temporal average $\bar{x}_i (t)$, estimated for each trajectory between $t$ and $\delta t$ (as defined in equation~\ref{gel:eq:temporal_variance}), the results will differ.

\begin{figure}[ht!]
        \centering
        \begin{subfigure}[b]{0.5\textwidth}
                \includegraphics[width=\textwidth]{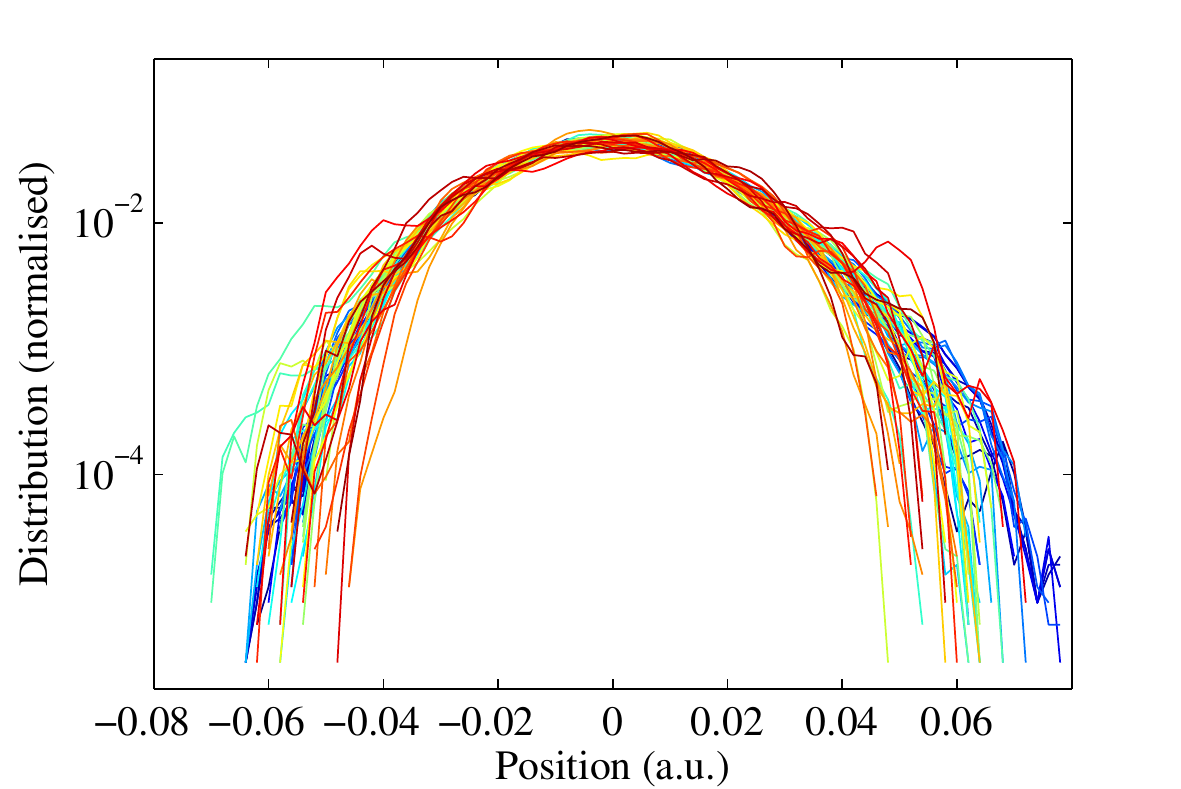}
                \caption{When one subtracts the ensemble average.}
                \label{gel:fig:PDF_ensemble}
        \end{subfigure}%
        \begin{subfigure}[b]{0.5\textwidth}
                \includegraphics[width=\textwidth]{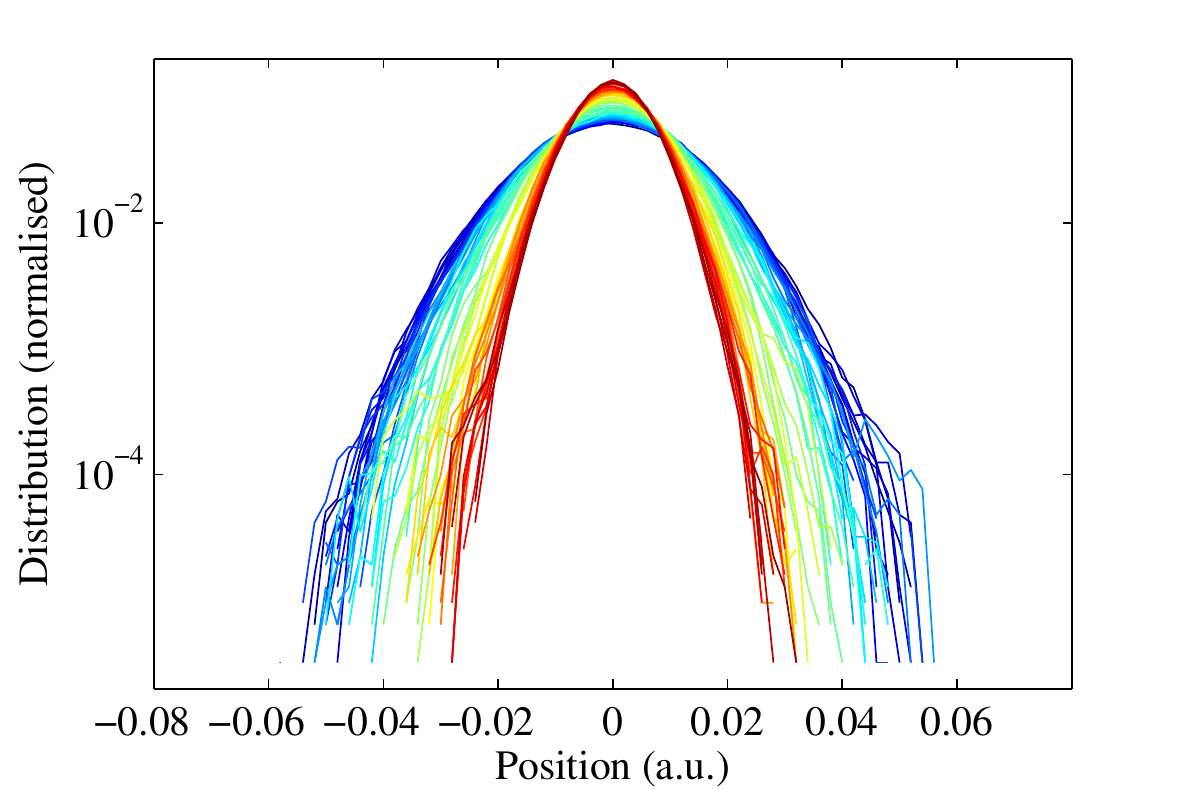}
                \caption{When one subtracts the temporal average.}
                \label{gel:fig:PDF_time}
        \end{subfigure}
        \caption{Evolution of the Probability Distribution Function of the position fluctuation $\delta x$ depending on the definition taken for the subtracted average. The PDFs are computed on a time-window $\delta t = \SI{0.5}{\second}$ for different times after the quench, going from $t = \SI{0}{\second}$ (blue curves) to $t = \SI{540}{\second}$ (red curves).}\label{gel:fig:PDFs}
\end{figure}

\begin{figure}[ht!]
\begin{center}
\includegraphics[width=9cm]{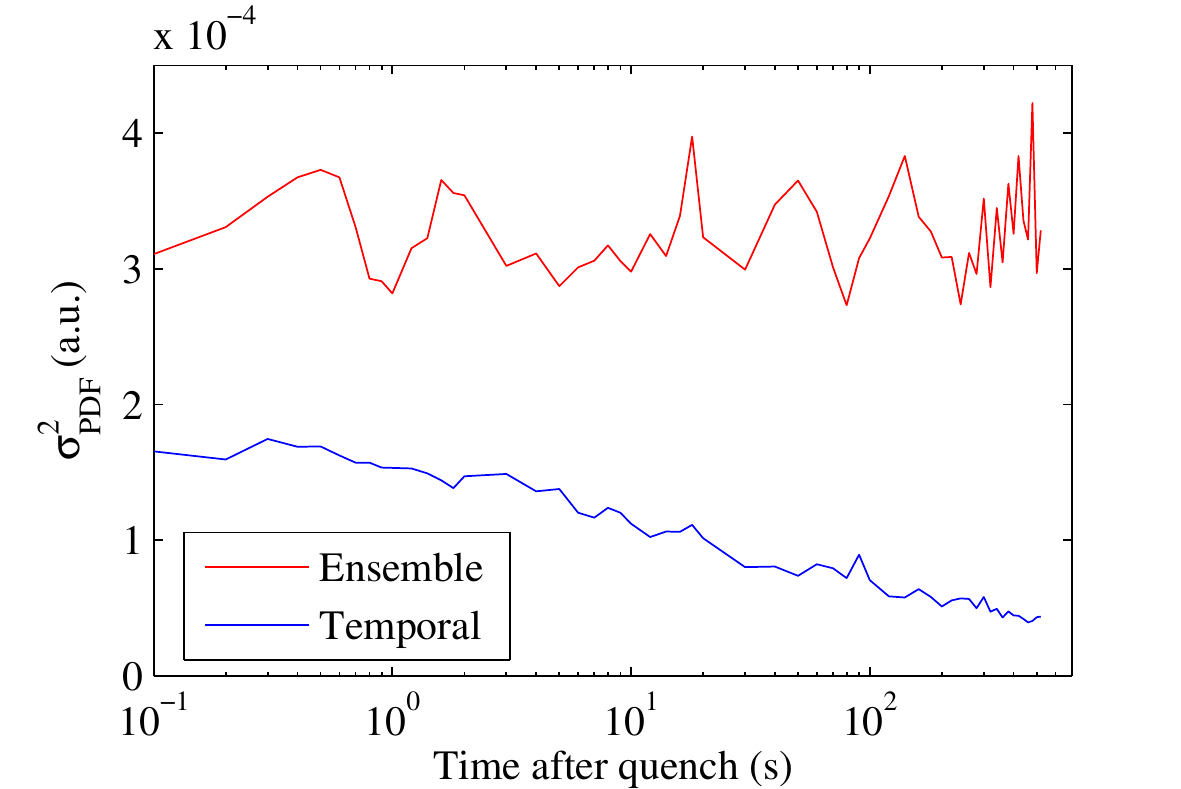}
\caption{Evolution of the variance estimated by fitting the PDFs with a Gaussian, at different times after the fast quench. When subtracting the correct ensemble average the variance is constant. When subtracting the temporal average, the variance decreases almost logarithmically with the time after the quench.} \label{gel:fig:var_PDF}
\end{center}
\end{figure}

As an example, we take the data of 132 quenches at \SI{27.5}{\celsius}, sampled at \SI{8}{\kilo\hertz}. The parameters are: melting time $\tau_{\text{melt}} = \SI{200}{\second}$, melting intensity $I_{\text{melt}} = \SI{245}{\watt}$, resting time $\tau_{\text{rest}}~=~\SI{570}{\second}$, and trapping intensity $I_{\text{trap} = \SI{26}{\watt}}$ which corresponds to trap stiffness\footnote{The trap stiffness was measured in water (where viscosity is known) for the same laser intensity.} $k \sim \SI{5}{\stiffness}$.
We compare the Probability Distribution Function of the positions with a $\delta t = \SI{0.5}{\second}$ at different times after the quench, when we subtract either the ensemble average (figure~\ref{gel:fig:PDF_ensemble}) or the temporal average (figure~\ref{gel:fig:PDF_time}). In the first case, the PDFs are nearly always Gaussian and do not evolve in time. In the second case, the PDFs are always nice gaussians, but with a variance that decreases in time. This is consistent with the previous results showing that the ensemble variance is constant at any time after the quench, whereas the temporal variance decreases logarithmically with the time after the quench. And the variances estimated by doing a Gaussian fit on the PDFs clearly shows the same behaviour (see figure~\ref{gel:fig:var_PDF}).

This effect is simple to understand: the trajectories evolve on a time $\tau_{\text{gel}}$. This time is much bigger than $\tau_{\text{fluc}}$, the typical time of the fluctuations, and $\delta t$. On the time-window $\delta t$, each portion of trajectory $x_i(t)$ can be written $x_i (t) = \bar{x}_i + \delta x_i (t)$, where $\bar{x}_i$ is the time average of the $i^\text{th}$ trajectory over the time-window. When one considers the N trajectory fragments between $t$ and $t+\delta t$, the difference between them is mostly due to the averaged value $\bar{x}_i$ of each trajectory fragment, and not to the fast fluctuations $\delta x_i (t)$. Which means that the distribution of all the $x_i(t)$ between $t$ and $t+\delta t$ is nearly the same as the ensemble distribution of the $\bar{x}_i$. Whereas, the distribution of all the $\delta x_i (t)$ is nothing more than the distribution of the fast temporal fluctuations of one single trajectory.

This difference is very important, as any kind of high-pass filtering (for example a ``detrend'' function which is often used to suppress slow drifts) done to the trajectories will result in subtracting the temporal average, and thus distort the PDFs estimation.

The experimental results show that the correct estimated PDFs do not evolve in time after a fast quench. Since we have already shown that the correct ensemble variance always verifies the equilibrium equipartition relation, we can conclude that the variance of these PDFs is simply $k_{\text{B}}T/k$. It is again interesting to see that, even if the gelatin solution is aging, its ensemble statistical properties seem to verify relations that are normally verified at equilibrium.

It was also verified with some available data from~\cite{TheseRuben,RubenPRL2011,RubenEPL2012} that the correct ensemble PDFs are not evolving with time after the quench.

\section{What about heat and Fluctuation Dissipation Theorem?}

In previous works~\cite{TheseRuben,RubenPRL2011,RubenEPL2012} the anomalous fluctuations observed right after the quench were interpreted in terms of heat exchanges between the bath and the particle. Indeed, the heat exchanged between $t$ and $t+\tau$ is equal to the variation of the particle's energy $\Delta U_{t,\tau} = \Delta U_{t+\tau} - \Delta U_{t}$:
\begin{equation}
Q_{t,\tau} = \Delta U_{t,\tau} = \frac{k}{2} \left( x^2(t+\tau) - x^2(t) \right).
\end{equation}
In particular, the decrease of the variance after the quench has been interpreted as the sign of a heat transfer from the particle to the bath :
\begin{equation}
\langle Q_{t,\tau}  \rangle = \frac{k}{2} \left( \sigma^2(t+\tau) - \sigma^2(t) \right) \leq 0.
\end{equation}
The Probability Distribution Functions (PDF) of the $Q_{t,\tau}$ were shown to be asymmetrical for values of $t$ and $\tau$ chosen right after the quench (\textit{i.e.} where the anomalous fluctuations were observed).

A violation of Fluctuation-Dissipation Relation was also observed for times right after the fast quench. It was linked to the non-zero heat exchange by a modification of the Harada-Sasa equality~\cite{HaradaSasa2005,Verley2012} for non-stationary systems:
\begin{equation}
\int_{1/\Delta t}^{\infty} \left[ S_{x}(t,f) - \frac{2 k_{\text{B}} T}{\pi f} \text{Im} \{ \hat{R}(t,f) \} \right] \, \mathrm df = \frac{2 | \langle Q_{t,\Delta t}\rangle |}{k}
\end{equation}
Where $S_{x}(t,f)$ is the Power Spectral Density of $x$ and $\hat{R}(t,f)$ is the Fourier transform of the linear response function of the position $x$ to a perturbative time-dependent force (these two quantifies are function of the frequency $f$, but also of the time $t$ since the system is ageing).

All these interpretations comes from the fact that the variance was seen anomalously high right after the quench, and then reduces to the equipartition value after a given time. In particular, the asymmetry and the shape of the PDFs of $Q_{t,\tau}$ are simply mathematical consequences of the fact that $x(t+\tau)$ and $x(t)$ have Gaussian PDFs with different variances $\sigma^2(t+\tau) > \sigma^2(t)$. For example, the exchange Fluctuation Theorem (xFT) that was retrieved with the asymmetrical PDFs of $Q$ is mathematically verified for any random variable defined by $y = x_1 - x_2$ where $x_1$ and $x_2$ are random variables with centered Gaussian distribution of different variances: $\sigma_{x_1}^2 \neq \sigma_{x_2}^2$. 

Since we have already shown that, if estimated correctly, the PDFs of $x$ show no anomalous behavior and have a constant variance equal to $k_{\text{B}}T/k$, if follows directly that the PDFs of $Q_{t,\tau}$ are symmetrical. Consequently, in average there is no heat exchange between the particle and the bath, for any $t$ and $t+\tau$, and the xFT reduces to a trivial equality because the asymmetry function of a symmetrical distribution is always zero.

Considering the Fluctuation-Dissipation Theorem, one must remind that it is \textit{a priori} not a good idea to test it in Fourier space. Indeed it is necessary to assume that the system is stationary and ergodic to link the correlation function to the power spectrum with the Wiener–Khinchin theorem~\cite{Wiener,Khintchine}. Therefore, when the system is not stationary, one should in theory look at the proper ensemble correlation function:
\begin{equation}
\text{EnsCorr}_{xx}(t,\tau) =  \frac{1}{N} \sum_{i=1}^{N} \left[ x_{i}(t)-\langle x (t) \rangle \right] \times \left[ x_{i}(t+\tau)-\langle x (t+\tau) \rangle \right]
\end{equation}
instead of the Power Spectral Density (PSD), which is a temporal quantity. Of course, one can always define a PSD of $x_{i}$ on a given time-window $\delta t$ for each trajectory $S_{x_{i}}(t,f)$. And this PSD would be equal to the Fourier Transform (FT) of the temporal correlation of $x_{i}$ computed on the same time-window:
\begin{equation}
\text{TimeCorr}_{xx}(t,\tau) = \frac{1}{\delta t} \int_{t}^{t+\delta t} \left[ x_{i}(t^{\prime})-\bar{x_{i}} \right] \times \left[ x_{i}(t^{\prime}+\tau) - \bar{x_{i}} \right] \, \mathrm dt^{\prime} = \mathrm{FT}\{ S_{x_{i}}(t,f) \}.
\end{equation}
But the system needs to be considered stationary and ergodic on the time-window $\delta t$, so that the ensemble and temporal correlations should be equal.

Here, the assumption of local stationarity is reasonable since the PSD were computed on \SI{15}{\second} long time-windows (which is short compared to the $\sim \SI{900}{\second}$ necessary to gel). However, it seems probable that the observed violation of Fluctuation-Dissipation Theorem was only due to the same kind of artifact already responsible for anomalous variance increase (for example: slow drifts for times right after the fast quench), because PSDs are sensible to low-frequency noises. Thus, there is no reason that this apparent violation is linked to an heat exchange, which does not exist anyway.

\begin{figure}[ht!]
\begin{center}
\includegraphics[width=11cm]{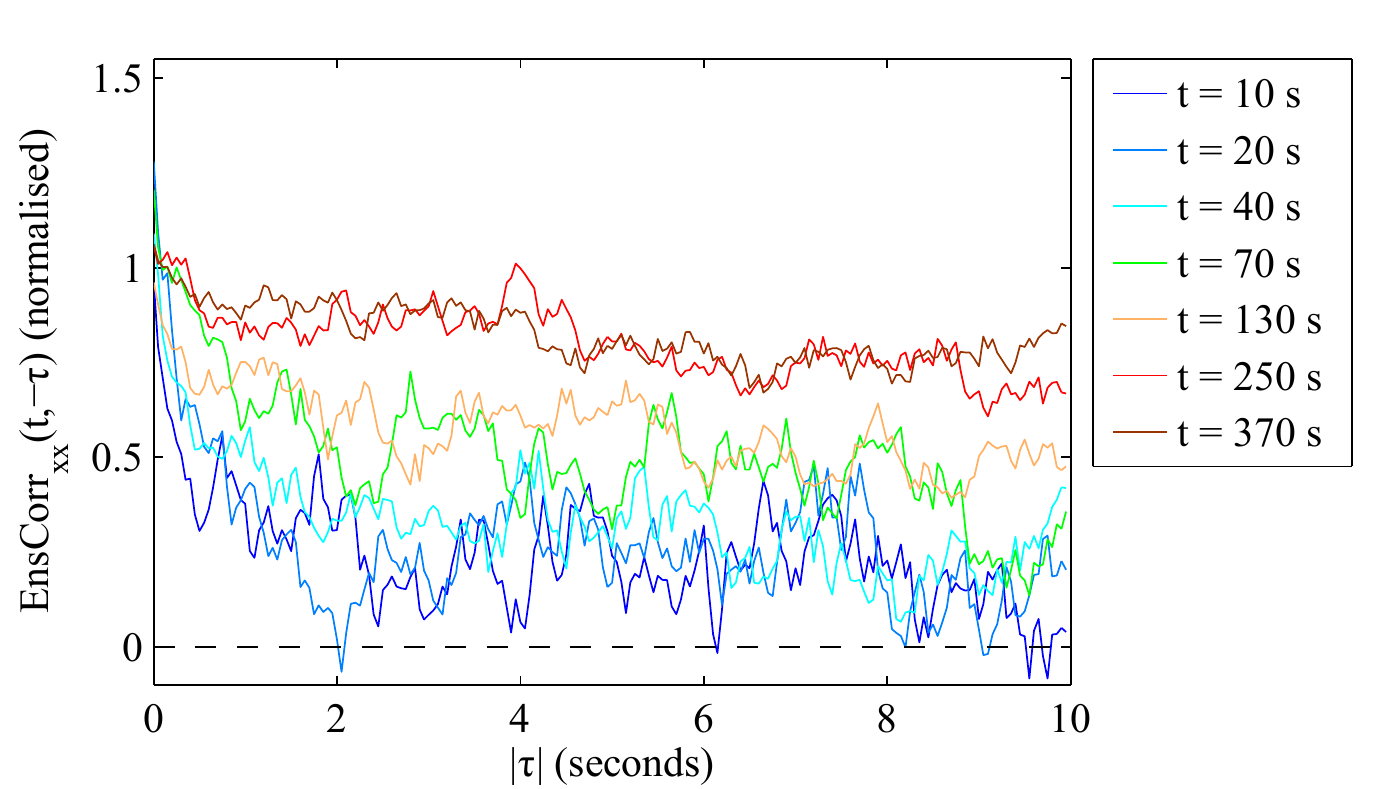}
\caption{Normalised ensemble correlation function for a quench at \SI{27}{\celsius}. Here we keep $t$ fixed and we vary $\tau$ from \SI{-10}{\second} to \SI{0}{\second}. The normalisation is done by dividing $\text{EnsCorr}_{xx}(t,\tau)$ by the value of $k_{\text{B}} T/k$ extracted from the variance of the position PDFs.} \label{gel:fig:corr_different_t}
\end{center}
\end{figure}

Some ensemble correlation functions of the particle's position are shown in figure~\ref{gel:fig:corr_different_t} for a set of 40 quenches at \SI{27}{\celsius}, sampled at \SI{8}{\kilo\hertz}. The parameters are: melting time $\tau_{\text{melt}} = \SI{200}{\second}$, melting intensity $I_{\text{melt}} = \SI{270}{\watt}$, resting time $\tau_{\text{rest}}~=~\SI{570}{\second}$, and trapping intensity $I_{\text{trap} = \SI{26}{\watt}}$ which corresponds to trap stiffness\footnote{The trap stiffness was measured in water (where viscosity is known) for the same laser intensity.} $k \sim \SI{5}{\stiffness}$.
The data are very noisy, but there is a tendency: the characteristic time increases after the quench, which is reasonable since the gelatin viscosity is also increasing during the gelation. The correlation functions are not simply exponential relaxations, which is consistent with the fact that the PSD are not Lorentzian (as shown in figure~\ref{gel:fig:spectrum_320min}). 

\begin{figure}[ht!]
\begin{center}
\includegraphics[width=9cm]{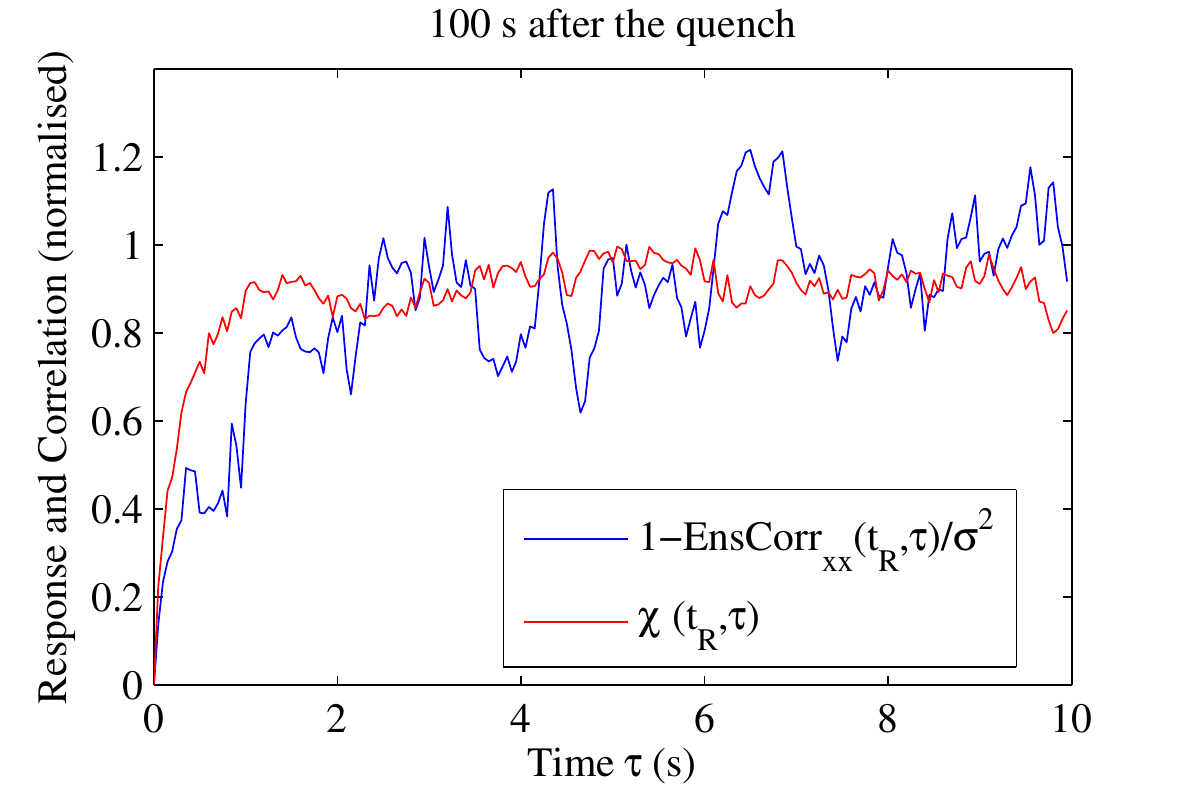}
\caption{Normalised response function $\chi(t_{R},\tau)$ and ensemble correlation function for $t_{R}~=~\SI{100}{\second}$ after the quench, and $\tau$ going from $0$ to $\SI{10}{\second}$.} \label{gel:fig:verif_FDT}
\end{center}
\end{figure}

We also made some experimental tests of Fluctuation Dissipation Theorem (FDT), by looking at the ensemble correlation of $x$ and the response to an Heaviside change of the position of the trap. For these measurements, the position of the trap is changed from $X_1$ to $X_2$ at a time $t_{R}$ after the first quench, and  the sample is let gel with the particle  in $X_2$. Then, for the second quench, the position of the trap is moved back to $X_1$ at time $t_{R}$ after the quench, and the sample is let gel in $X_1$. The procedure is then repeated alternatively.
The perturbation introduced by the change of trapping position allows us to compute a normalized response function, averaged over the trajectories:
\begin{equation}
\chi (t_{R},\tau) = \frac{\left\langle x(t_{R}+\tau) - X_{\text{initial}} \right\rangle}{X_{\text{final}} - X_{\text{initial}}}
\end{equation}
where $[X_{\text{initial}};X_{\text{final}}] = [X_1;X_2]$ or $[X_2;X_1]$.
It corresponds to the usual definition of the response function:
\begin{equation}
\chi (t) = \frac{\left\langle x(t)_{\text{perturbed}}-x(t)_{\text{unperturbed}} \right\rangle}{\text{perturbation amplitude}}.
\end{equation}
We use $X_{\text{final}} - X_{\text{initial}}$ which is proportional to the perturbation amplitude. And we simply take $X_{\text{initial}}$ as the average value of the unperturbed trajectory, because the mean position of the bead is constant and equal to the position of the trap if there is no perturbation\footnote{One could also take $\langle x (t_{R}) \rangle$ to guarantee that $\chi (t_{R},0) = 0$, but it wasn't necessary here.}.\\
If the FDT is verified, the response function should verify:
\begin{equation}
\chi (t_{R},\tau) = 1 - \frac{k}{k_{\text{B}} T} \text{EnsCorr}_{xx}(t_{R},\tau)
\end{equation} 
Some data are presented in figure~\ref{gel:fig:verif_FDT} for 50 quenches at \SI{26}{\celsius}, sampled at \SI{8}{\kilo\hertz}. The parameters are: melting time $\tau_{\text{melt}} = \SI{200}{\second}$, melting intensity $I_{\text{melt}} = \SI{270}{\watt}$, resting time $\tau_{\text{rest}}~=~\SI{570}{\second}$, and trapping intensity $I_{\text{trap} = \SI{26}{\watt}}$ which corresponds to trap stiffness\footnote{The trap stiffness was measured in water (where viscosity is known) for the same laser intensity.} $k \sim \SI{5}{\stiffness}$.
The values of $X_1$ and $X_2$ are estimated by computing the mean position of the bead when the gelatin is melted (which gives alternatively $X_1$ and $X_2$). The exact value of $k_{\text{B}} T/k$ was extracted from the variance of the position PDFs computed before changing the position of the trap. These measurements are a bit noisy because it requires a lot of statistics to compute a proper ensemble correlation function, but no apparent violation of the FDT was found for the times tested.

We didn't test the Fluctuation Dissipation Theorem for times $t_{R}$ taken shortly after the quench, because the ensemble correlation shows a characteristic time which is very short at this time (see figure~\ref{gel:fig:corr_different_t}). It is then more difficult to compute a proper ensemble correlation right after the quench, than when the viscosity of gelatin has already started to increase. We also didn't compute the response function by varying $t_{R}$ for a fixed $t_{R} + \tau$, because it would require a lot of time to do the experiments. Indeed, each set of $t_{R}$ requires one day of measurement to compute $\chi (t_{R},\tau)$, and the sample cannot be kept a lot of days without degrading.

\newpage

\section{Conclusion}

In conclusion, we have locally studied the gel transition of a gelatin solution. We were unable to reproduce the results of previous works~\cite{TheseRuben,RubenPRL2011,RubenEPL2012}, but we have identified some experimental and data analysis artifacts which may explain the effects previously observed. In particular we have analyzed the effect of time-windows on proper ensemble averages, which are important to study aging systems.

We have shown that in the hysteresis range of temperature ($\SI{28.3}{\celsius} < T < \SI{36}{\celsius}$), bulk gelation can occur on very long times, and viscoelastic properties gradually appear. The characteristic time of the particle trapped in the bulk-gelling sample was seen to decrease exponentially before the gelation (whereas the viscosity evolves logarithmically after the gelation).

For fast quenches of a small droplet of gelatin solution, we have found that the Probability Distribution Functions of the position of the trapped particle do not evolve with time after the quench, even if the gelatin sample is undergoing aging and the viscoelastic properties are clearly evolving. Moreover, these PDFs show equilibrium-like properties, being Gaussian with a variance equal to the equipartition value $k_{\text{B}} T/k$. These results seem not so surprising \textit{a posteriori}, since it was already observed in the previous works that, after $\sim \SI{15}{\second}$ the Brownian motion of the trapped particle behaves like in equilibrium with the thermal motion of the gelatin chains. Only the very first seconds after the quench showed anomalous behavior, which was strange, because the complete gelation occurs on much larger scales ($\sim \SI{900}{\second}$). It however remains striking that a system which is clearly not stationary because of aging has ensemble properties which are stationary.

For systems which are not ergodic or not stationary, time properties can be very different from ensemble properties. And it was also shown that some artifacts (like slow drifts) or analysis bias (like high-pass filter) can greatly modify the results if ensemble properties are estimated on time-windows. Therefore, one must be very careful when studying statistical properties of an aging system. This kind of problems had already arisen for other aging systems. For example, it was already shown in~\cite{Jop2009} that increase in effective temperature previously seen in suspension of Laponite~\cite{Greinert2006} were in fact artifacts due to analysis methods.

Finally, in agreement with the absence of anomalous position fluctuation after the fast quench, no heat exchange, nor violation of the Fluctuation Dissipation Theorem was seen, as it would be expected in an equilibrium medium.

\bibliographystyle{ieeetr}
\bibliography{biblio}

\end{document}